\documentclass[12pt]{article}
\usepackage{amsfonts,amsmath,amssymb,epsf,cite}
\usepackage{graphicx,color}
\usepackage[dvipdfm]{hyperref}
\renewcommand{\baselinestretch}{1.1}

\def\d#1{\,{\rm d}#1}

 \def\ep{{\epsilon}}
 \def\d{{\delta}}
 
 \def\t{{\theta}}

 \def\frac#1#2{{#1\over #2}}

 \def\D{{\Delta}}
 \def\g{{\gamma}}
 \def\s{\sqrt}
 
 \def\b{{\beta}}

 \def\CN{{\cal N}}
 \def\p{\partial}

 \def\de{\partial}

 \def\f {\frac}
 \def\ti{\tilde}
 \def\ap{\alpha}

 \def\ddd{\cdot\cdot\cdot}
 \def\no{\nonumber \\}
 \def\la{\langle}
 \def\lb{\rangle}
 \def\ep{\epsilon}

 \def\vp{\varphi}

 \def\ep{{\epsilon}}
 \def\d{{\delta}}
 
 \def\t{{\theta}}

 \def\frac#1#2{{#1\over #2}}

 \def\D{{\Delta}}
 \def\g{{\gamma}}
 \def\s{\sqrt}
 
 \def\b{{\beta}}

 \def\CN{{\cal N}}
 \def\p{\partial}

\def\be{\begin{equation}}
\def\ee{\end{equation}}
\def\ba{\begin{eqnarray}}
\def\ea{\end{eqnarray}}

\begin{document}
\begin{titlepage}
\thispagestyle{empty}
\begin{flushright}
KUNS-2207\\
IPMU09-0056
\end{flushright}

\bigskip

\begin{center}
\noindent{\Large \textbf{Holographic Entanglement Entropy: An Overview}}\\
\vspace{2cm} \noindent{Tatsuma Nishioka$^{a,b}$\footnote{e-mail:nishioka@gauge.scphys.kyoto-u.ac.jp},
Shinsei Ryu$^{c}$\footnote{e-mail:sryu@berkeley.edu}
and Tadashi Takayanagi$^{b}$\footnote{e-mail:tadashi.takayanagi@ipmu.jp}}

\vspace{1cm}
  {\it
 $^{a}$Department of Physics, Kyoto University, Kyoto, 606-8502, Japan\\
 $^{b}$Institute for the Physics and Mathematics of the Universe, \\
 University of Tokyo, Kashiwa, Chiba 277-8582, Japan\\
  $^{c}$Department of Physics, University of California, Berkeley, CA 94720, USA}
\end{center}

\begin{abstract}
In this article, we review recent progresses on the holographic
understandings of the
entanglement entropy in the AdS/CFT correspondence.
After reviewing the general idea of holographic entanglement entropy, we will
explain its applications to confinement/deconfinement phase transitions, black hole
entropy and covariant formulation of holography.

\end{abstract}

\end{titlepage}

\newpage

\tableofcontents

\newpage

\section{Introduction}
\label{intro} \setcounter{equation}{0}
\hspace{5mm}
In recent developments of string theory, the idea of holography has obviously
played crucial roles. Holography claims that the degrees of freedom in $(d+2)$-dimensional
quantum gravity are much more reduced than we naively think, and
will be comparable to those of quantum many body systems in $d+1$
dimensions \cite{holography,BiSu}. This was essentially found by remembering that
the entropy of a black hole
is not proportional to its volume, but to its area of the
event horizon $\Sigma$ (the Bekenstein-Hawking formula \cite{BeHa}):
\be
S_{BH}=\f{\mbox{Area}(\Sigma)}{4G_N}\ ,\label{BHF}
\ee
where $G_N$ is the Newton constant.
Owing to the discovery of the AdS/CFT correspondence
\cite{Maldacena}, we know explicit examples where the holography
is manifestly realized. The AdS/CFT argues that the quantum gravity on $(d+2)$-dimensional
anti-de Sitter spacetime (AdS$_{d+2}$) is equivalent to a certain conformal field
theory in $d+1$ dimensions (CFT$_{d+1}$) \cite{Maldacena,ADSGKP,ADSWitten,adsreview}.

Even after quite active researches of AdS/CFT for these ten years, fundamental mechanism of
the AdS/CFT correspondence still remains a mystery, in spite of so many of evidences in various examples.
In particular, we cannot answer which region of AdS is responsible to particular information
in the dual CFT. To make modest progresses
for this long standing problem, we believe that
it is important to understand and formulate the holography in terms of
a universal observable,
rather than quantities which depend on the details of theories such as specific operators or
Wilson loops etc. We only expect that a quantum gravity in some spacetime is dual to
(i.e. equivalent to) a certain theory which is governed by the law of quantum mechanics.
We would like to propose
that an appropriate quantity which can be useful in this universal
viewpoint is the entanglement entropy. Indeed, we can always define the entanglement entropy in
any quantum mechanical system.

The entanglement entropy $S_A$ in quantum field theories or quantum many body systems is a non-local
quantity as opposed to
correlation functions. It is defined
as the von Neumann entropy $S_A$ of the reduced density matrix
when we `trace out' (or smear out)
degrees of freedom inside a $d$-dimensional space-like submanifold
$B$ in a given $(d+1)$-dimensional QFT, which is a complement of $A$.
$S_A$ measures how the subsystems $A$ and $B$ are correlated with each other.
Intuitively we can also say that
this is the entropy for an observer in $A$ who is not accessible to
$B$ as the information is lost by the smearing out in region $B$.
This origin of entropy looks analogous to the black hole entropy.
Indeed, this was the historical motivation of considering the entanglement
entropy in quantum field theories \cite{Thooft,Bombelli,Srednicki}.
Interestingly, the leading divergence of $S_A$ is proportional
to the area of the subsystem $A$, called the area law
\cite{Bombelli,Srednicki} (refer also
to the review articles
\cite{Calabrese:2005zw,Riera:2006vj,Latorre:2006py,Amico,ECP,Casini:2009sr,CReview,LatR}).

Since $S_A$ is defined as a von Neumann entropy,
 we expect that the entanglement entropy is
directly related to the degrees of freedom. Indeed, in two-dimensional
conformal field theory, the entanglement
entropy is proportional to the central charge in two-dimensional
conformal field theories (2D CFTs) as shown in
\cite{HLW, Cardy}, where a general prescription of computing the
entropy in 2D CFTs is given.
 Also in the mass perturbed CFTs (massive QFTs) the same
conclusion holds \cite{Vidal03, Latorre04, Cardy}. Furthermore,
our holographic result shows that the similar statement is also true in
four or higher even-dimensional CFTs. As opposed to the thermal entropy,
the entanglement entropy is non-vanishing at zero temperature.
Therefore we can employ it to probe the quantum properties of the
ground state for a given quantum system. It is also a useful order parameter
of quantum phase transition at zero temperature as will be explained in
 Sec.\ \ref{secpha}.

Now we come back to our original question where in AdS
given information in CFT is saved.
Since the information included in a subsystem $B$ is evaluated by the entanglement
entropy $S_A$, we can formulate this question more concretely as follows: ``Which part of
AdS space is responsible for the calculation of $S_A$ in the dual gravity side ?''
Two of the authors of this article proposed a holographic formula of the entanglement entropy
in \cite{RuTa,RuTaL}:
\be
S_A=\f{\mbox{Area}(\gamma_A)}{4G^{(d+2)}_N}\ , \label{RTF}
\ee
where $\gamma_A$ is the $d$-dimensional minimal surface $\gamma_A$ whose boundary is given
by the $(d-1)$-dimensional manifold
$\de \gamma_A=\de A$
(see Fig.\ \ref{Fig:holo}); the constant $G^{(d+2)}_N$ is
the Newton constant of the general gravity in AdS$_{d+2}$.
This formula can be applied equally well to asymptotically
AdS static spacetimes. Originally, this formula (\ref{RTF}) is speculated from
the Bekenstein-Hawking formula (\ref{BHF}). Indeed, since the minimal surface
tends to wrap the horizon in the presence of event horizon, our formula (\ref{RTF}) can be regarded as a
generalization of the well-known formula (\ref{BHF}).
Also the area law of $S_A$ \cite{Bombelli,Srednicki} can
be automatically derived from our holographic description.

The purpose of this article is to explain this holographic description and then
to review its current status with recent progresses and applications
\cite{Emparan:2006ni}-\cite{ALT}.
In AdS$_3/$CFT$_2$, we can confirm that the formula
(\ref{RTF}) is precisely true by comparing the holographic result with the known
2D CFT results
\cite{RuTa,RuTaL}.
In higher-dimensional cases, however, the proposed formula
 (\ref{RTF}) has not been derived rigorously from the bulk to boundary relation in AdS/CFT
 \cite{ADSGKP, ADSWitten} at present.
 Also the direct calculations of the entanglement entropy in the CFT side is very complicated in higher dimensions.
 Nevertheless,
a heuristic derivation has been presented in \cite{Fursaev:2006ih} and many evidences
\cite{RuTaL,Solodukhin:2006xv, Hirata:2006jx, Nishioka:2006gr, Headrick:2007km, Solodukhin:2008dh, Casini:2008as}
have been found. Our holographic formula has also been successfully applied to the explanation of black hole
and de-Sitter entropy \cite{Emparan:2006ni, Iwashita:2006zj, Solodukhin:2006xv, Azeyanagi:2007bj}
(see also \cite{HMS,MBH} for earlier pioneering ideas on the entanglement entropy in AdS/CFT with
event horizon; see also \cite{Einhorn}), and
to an order parameter of a confinement/deconfinement phase transition
\cite{Nishioka:2006gr, Klebanov:2007ws, Faraggi:2007fu, Buividovich:2008kq, Fujita:2008zv, Buividovich:2008gq, Bah:2008cj, Buividovich:2008yv}.

In condensed matter physics,
the entanglement entropy is expected to be
a key quantity to understand
several aspects of quantum many-body physics.
A central question in quantum many-body physics is
how we can characterize different phases
and phase transitions.
While microscopic Hamiltonians in condensed matter systems
(electronic systems, in particular) are quantum mechanical,
a wide range of quantum phases turn out to have a classical analogue,
and if so, they can be understood in terms of symmetry breaking of some kind,
and in terms of classical order parameters.
On the other hand,
this paradigm, known from Landau and Ginzburg, does not always apply
when phases of our interest are inherently quantum.
Indeed, one of main foci in modern
condensed matter physics is to understand
quantum phases of matter
and
phase transitions between them,
which are beyond the Landau-Ginzburg paradigm.
To name a few, relatively well-understood examples,
the fractional quantum Hall effect,
and quantum magnets on some geometrically frustrated lattices
have attracted a lot of interest.
Many-body wavefunctions of quantum ground states in these phases look
featureless when one looks at correlation functions of local operators;
They cannot be characterized by classical order
parameters of some kind. Indeed, they should be distinguished by
their pattern of entanglement rather than their pattern of symmetry
breaking \cite{Wen89}. Thus, the entanglement entropy is potentially
useful to characterize these exotic phases\footnote{Recently, there have been a number of
progresses on holographic descriptions of various phase transitions analogous to
the ones in condensed matter physics, e.g. refer to the review \cite{HartR}}.

One can ask these questions
from a slightly more practical,
but ultimately fundamental,
point of view;
how can we simulate quantum states of matter efficiently
by classical computers?
The total dimension of the Hilbert space increases exponentially
as we increase the system size,
and hence bruteforce approaches
(e.g., exact diagonalization)
to quantum many-body systems are destined to fail.
It turns out having a good understanding on
how local regions of the whole quantum system are entangled to each other
would help to find good algorithms for quantum many-body problems,
such as the density matrix renormalization group (DMRG)\cite{DMRG}.
To be more precise,
the scaling of the entanglement entropy as a function of the size of
a given subregion of the system of interest
gives us a criterion for efficient approximability.
In other word, the entanglement entropy tells us
amount of information and degrees of freedom necessary to
represent a quantum ground state efficiently.

Reversing the logic, one can distinguish different phases of
quantum matter according to their computational complexity
and hence from the scaling of the entanglement entropy.
After all, what makes simulation of quantum systems by classical computers
difficult is nothing but entanglement.
Indeed, this idea has been pushed extensively in recent couple of
years for several 1D quantum systems. It has been revealed that
several quantum phases in 1D spin chains
can be distinguished by different scaling of the entanglement
entropy. See, for example,
\cite{Vidal03, Latorre04, Peschel04, Jin03, Its05} and
references in \cite{Cardy}.

For higher-dimensional condensed matter systems,
there have been many recent attempts in this direction.
In particular,
the entanglement entropy was applied
for so-called topological phases
in 2+1 dimensions
\cite{Kitaev05, Levin05}.
Typically, these phases
have a finite gap and
are accompanied
by many exotic features such as
fractionalization of quantum numbers,
non-Abelian statistics of quasi-particles,
topological degeneracy, etc.
They can be also useful
for fault tolerant quantum computations.
On the other hand, unconventional quantum liquid phases with gapless
excitations, such as gapless spin liquid phases, seem to be, at
least at present, more difficult to characterize in higher
dimensions. Our results from the AdS/CFT correspondence can be useful to
study these gapless (spin liquid) states (some of these phases have
been suspected to be described by a relativistic gauge field theory
of some sort \cite{Wen89}).

The organization of this paper is as follows:
In Sec.\ \ref{basics},
we go through some basic properties of
the entanglement entropy.
In particular,
we discuss how the entanglement entropy scales as a function
of the size of the subsystem in quantum field theories
and many-body systems.
Sec.\ \ref{holographic} presents our basic formula of the holographic entanglement entropy via AdS/CFT.
Many results mentioned in Sec.\ \ref{basics} are reproduced
from the holographic point of view.
In Sec.\ \ref{secpha},
we apply the entanglement entropy
as a non-local order parameter to
confinement/deconfinement phase transitions.
Sec.\ \ref{bfrs} reviews
two connections between
the entanglement entropy and the black hole entropy
obtained from the holographic calculation of the entanglement entropy.
In Sec.\ \ref{seccov}, we will explain a covariant formulation of
the holographic entanglement entropy. We conclude in Sec.\ \ref{seccon}
with a summary
and with possible future directions.


\section{Basics of Entanglement Entropy}
\label{basics}
\setcounter{equation}{0}
\hspace{5mm}
We start with a review of the definition and properties of the entanglement entropy.

\subsection{Definition of Entanglement Entropy}
\label{def EE}
\hspace{5mm} Consider  a quantum mechanical system with many degrees
of freedom such as spin chains. More generally, we can consider
arbitrary lattice models or quantum field theories (QFTs).
We put such a system at zero temperature and
then the total quantum system is described by the
pure ground state $|\Psi\lb$. We assume no degeneracy of the ground state.
Then, the density
matrix is that of the pure state \be \rho_{tot}=|\Psi\rangle \langle
\Psi| \ . \label{pure}\ee The von Neumann entropy of the total system
is clearly zero $ S_{tot}= -\mathrm{tr}\, \rho_{tot} \log
\rho_{tot}=0$.

\begin{figure}
\begin{center}
\includegraphics[width=10cm,clip]{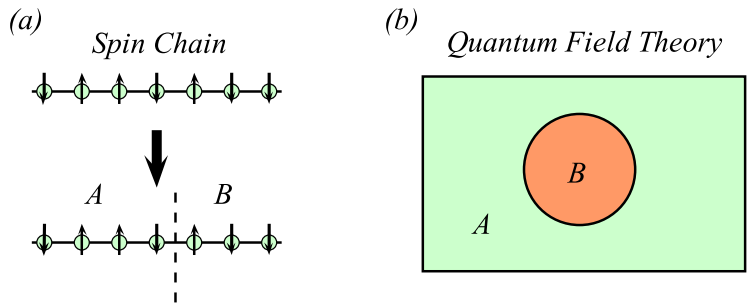}
\end{center}
\caption{
\label{fig:spinchain}
Examples of bipartitioning for the entanglement entropy.
A choice of the subsystems $A$ and $B$ is shown
for each of the two examples: $(a)$ a spin chain, $(b)$ a quantum field theory.
}
\end{figure}

Next we divide the total system into two subsystems $A$ and $B$
(see Fig.\ \ref{fig:spinchain}). In
the spin chain example,
we just artificially cut off the chain at
some point and divide the lattice points into two groups. Notice
that physically we do not do anything to the system and the cutting
procedure is an imaginary process. Accordingly the total Hilbert
space can be written as a direct product of two spaces
${\mathcal{H}}_{tot}={\mathcal{H}}_{A}\otimes {\mathcal{H}}_{B}$
corresponding to those of subsystems $A$ and $B$.
The observer who is only accessible to the subsystem $A$ will feel as
if the total
system is described by the reduced density matrix $\rho_A$
\be
\rho_A= \mathrm{tr}_{B}~\rho_{tot}\ ,
\ee where the
trace is taken only over the Hilbert space ${\mathcal{H}}_{B}$.

Now we define the entanglement entropy of the
subsystem $A$ as
the von Neumann entropy of the reduced
density matrix $\rho_A$
\begin{eqnarray}
S_A =
- \mathrm{tr}_{A}\,
\rho_{A} \log \rho_{A}\ .
\label{eq:def entanglement entropy}
\end{eqnarray}
This quantity provides us with a convenient
way to measure how closely entangled (or how ``quantum'') a given
wave function $|\Psi\rangle$ is.

In time-dependent
backgrounds the density matrices $\rho_{tot}$ and $\rho_A$ are time
dependent as dictated by the von Neumann equation. Thus we need to
specify the time $t=t_0$ when we measure the entropy. In this paper, we will always deal with
static systems except in Sec.\ \ref{seccov}.

It is also possible to define the entanglement entropy $S_A(\beta)$
at finite temperature $T=\beta^{-1}$.
This can be done just by replacing (\ref{pure}) with the thermal one
$\rho_{thermal}=e^{-\beta H}$,
where $H$ is the total Hamiltonian.
When $A$ is the total system, $S_A(\beta)$ is clearly
the same as the thermal entropy.
Also in general, if we take the high temperature limit $\beta\to 0$, then
the difference $S_{A_1}(\beta)-S_{A_2}(\beta)$ approaches the difference of thermal entropy between $A_1$ and
$A_2$. This subtraction is necessary to cancel the ultraviolet divergences as explained later.

\subsection{Properties}
\label{prop EE}
\hspace{5mm} There are several useful properties which
the entanglement entropy enjoys generally.
We summarize some of them as follows (the derivations and other properties of
the entanglement entropy can be found in e.g. the textbook \cite{Nielsen-Chuang00}):
\begin{itemize}
\item If the density matrix $\rho_{tot}$ is pure such as in the zero temperature system, then
we find the following relation assuming $B$ is the complement of $A$:
\be S_A=S_B\ . \label{ext}
              \ee
        This manifestly shows that the entanglement
        entropy is not an extensive quantity.
              This equality (\ref{ext}) is violated
              at finite temperature.

\item
For any three subsystems $A$, $B$ and $C$
that do not intersect each other,
the following inequalities hold :
\begin{eqnarray}
 S_{A+B+C} + S_{B} &\le& S_{A+B}+S_{B+C} \ ,  \label{Stronga}\\
 S_{A} + S_{C} &\le& S_{A+B}+S_{B+C} \ , \label{Strongb}
\end{eqnarray}
These inequalities are called the strong subadditivity \cite{LiRu}, which is the most powerful
inequality obtained so far with respect to the entanglement entropy.
In \cite{Casinicth,Casini:2006es} (see also \cite{Casiniarea}), the authors
presented an entropic proof of the c-theorem by applying the strong subadditivity to 2d quantum field theories.

\item By setting $B$ empty in (\ref{Stronga}), we can find the subadditivity relation
\be
 S_{A+B}\le S_{A}+S_{B} \ . \label{Sub}
\ee
The subadditivity (\ref{Sub}) allows us to define an interesting quantity called mutual information
$I(A,B)$ by
\be
I(A,B)=S_A+S_B-S_{A+B}\geq 0 \ .
\ee

\end{itemize}

\subsection{Entanglement Entropy in QFTs}
\label{EE in QFT, Area}
\hspace{5mm}  Consider a QFT on a $(d+1)$-dimensional manifold
$\mathbb{R}\times N$, where $\mathbb{R}$ and $N$
 denote the time direction and the $d$-dimensional
 space-like manifold, respectively.
We define the subsystem by a $d$-dimensional submanifold $A\subset
N$ at fixed time $t=t_0$. We call its complement the submanifold
$B$. The boundary of $A$, which is denoted by $\de A$, divides the
manifold $N$ into two submanifolds $A$ and $B$. Then we can define
the entanglement entropy $S_A$ by the previous formula (\ref{eq:def
entanglement entropy}). Sometimes this kind of entropy is called
geometric entropy as it depends on the geometry of the submanifold
$A$. Since the entanglement entropy is always divergent in a
continuum theory,
we introduce an
ultraviolet cut off $a$ (or a lattice spacing).
 Then the coefficient in
front of the divergence turns out to be proportional to the area of
the boundary $\de A$ of the subsystem $A$ as first pointed out in
\cite{Bombelli,Srednicki}, \be S_A=\gamma\cdot \f{\mbox{Area}(\de
A)}{a^{d-1}}+\mbox{subleading terms} \ ,
 \label{divarea}\ee
where $\gamma$ is a constant which depends on the system. This
behavior can be intuitively understood since the entanglement
between $A$ and $B$ occurs at the boundary $\de A$ most strongly.
This result (\ref{divarea}) was originally found from numerical
computations \cite{Srednicki,Bombelli} and checked in many later
arguments (see e.g. recent works \cite{Eisert, Das, Casiniarea} ).

The simple area law (\ref{divarea}), however, does not always
describe the scaling of the entanglement entropy in generic
situations. Indeed the
entanglement entropy of 2D CFT scales
logarithmically with respect to the length $l$ of $A$ \cite{HLW,Cardy}.
If we assume the total system is infinitely long,
it is given by the simple formula \cite{HLW,Cardy}
\be
S_A=\frac{c}{3}\log \f{l}{a} \ , \label{enttdcft}
\ee
where $c$ is the central charge of the CFT.
As we will see in Sec.\ \ref{holographic},
the scaling behavior (\ref{enttdcft})
is
consistent with the generic structure (\ref{genfa})
expected from AdS/CFT.

Other situations
such as a compactified circle at zero
temperature or an infinite system at finite temperature can be
treated by applying the
conformal map technique and analytic formulas have been
obtained in \cite{Cardy}.
The results are given as follows
\ba
&& S^{c.c.}_A = \f{c}{3}\cdot\log\left(\f{L}{\pi a}\sin\left(\f{\pi
l}{L}\right)\right) \ , \label{entropyone} \\
&& S^{f.t.}_A=\f{c}{3}\cdot\log\left(\f{\beta}{\pi
a}\sinh\left(\f{\pi l}{\beta}\right)\right) \ ,
\label{entropytemp}
\ea
respectively, where $L$ is the circumference of the circle.

The result for a finite size system at finite temperature has been obtained
in \cite{Azeyanagi:2007bj} for a free Dirac fermion (i.e. $c=1$) in two dimensions.
In the high temperature expansion, the result becomes (we set $L=1$)
\ba S_{A}(\beta,l)
\!\!&=&\!\!
\f{1}{3}\log\left[\f{\beta}{\pi
 a}\sinh\left(\f{\pi
l}{\beta}\right)\right]+\f{1}{3}\sum_{m=1}^\infty
\log\left[\f{(1-e^{2\pi \f{l}{\beta}}e^{-2\pi \f{m}{\beta}
})(1-e^{-2\pi \f{l}{\beta}}e^{-2\pi \f{m}{\beta}})}{(1-e^{-2\pi
\f{m}{\beta}})^2}\right]\no  &&\quad + 2\sum_{k=1}^\infty
\f{(-1)^{k}}{k}\cdot\f{\f{\pi kl}{\beta}\coth\left(\f{\pi
kl}{\beta}\right)-1}{\sinh\left(\pi \f{k}{\beta}\right)} \ .
\label{toth}\ea
Using this expression, we can find the relation between
the thermal entropy $S^{thermal}(\beta)$ and the entanglement entropy
\be
S^{thermal}(\beta)=\lim_{\ep\to 0}\Bigl(S(\beta,1-\ep)-S(\beta,\ep)\Bigr) \ .
\ee

For conformal field theories in higher dimensions ($d>1$),
our holographic method discussed in Sec.\ \ref{AdS d+2/CFT d+1}
predicts the following general form of $S_A$
for relativistic quantum field theories,
assuming that $\de A$ is a smooth and compact
manifold
\ba
S_A \!\!&=&\!\!
  p_1  \left(l/a\right)^{d-1}
+ p_3 \left(l/a \right)^{d-3}
+\cdots   \no
&&
\cdots +\left\{
\begin{array}{ll}
\displaystyle p_{d-1}\left(l/a\right) + p_d  \ , &
 \mbox{$d$: even}    \\
\displaystyle p_{d-2} \left(l/a\right)^{2} + \ti{c} \log
\left(l/a\right) \ ,
&  \mbox{$d$: odd}    \\
\end{array}
\right.\label{genfa}
\ea
where $l$ is the typical length scale of $\partial A$.
This result includes the known result for $d=1$ (\ref{enttdcft}) with $\tilde{c}=c/3$.
Also, in the case of (3+1)-dimensional
conformal field theories ($d=3$),
the scaling law (\ref{genfa})
has been confirmed by direct field theoretical calculations
based on Weyl anomaly \cite{RuTaL},
where again the coefficient of the logarithmic term $\tilde{c}$
is given in terms of central charges of $(3+1)$ CFTs.
For $d=\mathrm{even}$,
Eq.\ (\ref{genfa}) has been the only known analytical result
for (interacting) conformal field theories.
Even though we assumed conformal field theories in the above, the same
scaling formula (\ref{genfa}) should be true for a quantum field theory
with a UV fixed point, i.e.,
at a relativistic quantum critical point.

When the boundary $\de A$ is not a smooth manifold such as the one with cusp
singularities, we will have other terms $(l/a)^{d-2},(l/a)^{d-4},\ddd$
which do not obey the scaling law in (\ref{genfa}) \cite{Fu}. For example, for a three-dimensional
CFT ($d=2$), if $\de A$ has a cusp with the angle $\Omega$, then $S_A$ includes a
logarithmic term $\sim -f(\Omega)\log l/a$, for a certain function $f$
\cite{Casini:2006hu,Casini:2008as}. Refer also to
Sec.\ \ref{EE for cusps} for more details.

Also, if we consider a gapped system in three dimensions ($d=2$)
which is described (at low energies) by a topological field theory
(called topologically ordered phase), the scaling of the entanglement entropy
is the same as (\ref{genfa}) with $d=2$.
The constant $p_d=p_2$
in a topologically ordered phase is, however,
invariant under a smooth deformation of the boundary $\de A$. In this
case, $S_{top}=p_2$ is called the topological entanglement entropy \cite{Levin05,Kitaev05}.

It has been also pointed
out that the area law is corrected by a logarithmic factor as $S_A
\propto  (l/a)^{d-1}\log l/a+\ $(subleading terms) for fermionic
systems in the presence of a finite Fermi surface
(where, again, $l$ is the
characteristic length scale of the $(d-1)$-dimensional manifold
$\de A$) \cite{Wolf05, Gioev05, Barthel06, Li06, FaZi}.

\subsection{How to Compute Entanglement Entropy in QFTs}
\label{how to EE}
\hspace{5mm}
It is helpful to know how to calculate the entanglement entropy generally
in QFTs for later arguments. We will follow the method considered in \cite{Cardy}.
For this, we first
evaluate $\mathrm{tr}_A\,\rho_A^n$, differentiate it with
respect to $n$ and finally take the limit $n\to 1$ (remember that
$\rho_A$ is normalized such that $\mathrm{tr}_A\,\rho_A=1$)
\begin{eqnarray}
S_{A}
&=&
-\frac{\partial}{\partial n}
\mathrm{tr}_A\,\rho_A^n|_{n=1}
=
-\f{\de}{\de n}\log \mathrm{tr}_A~\rho_A^n|_{n=1} \ .
\label{deri}
\end{eqnarray}
This is called the replica trick. Therefore, what we have to do is
to evaluate $\mathrm{tr}_A~\rho_A^n$ in a given QFT.

\begin{figure}
\begin{center}
\includegraphics[height=8cm]{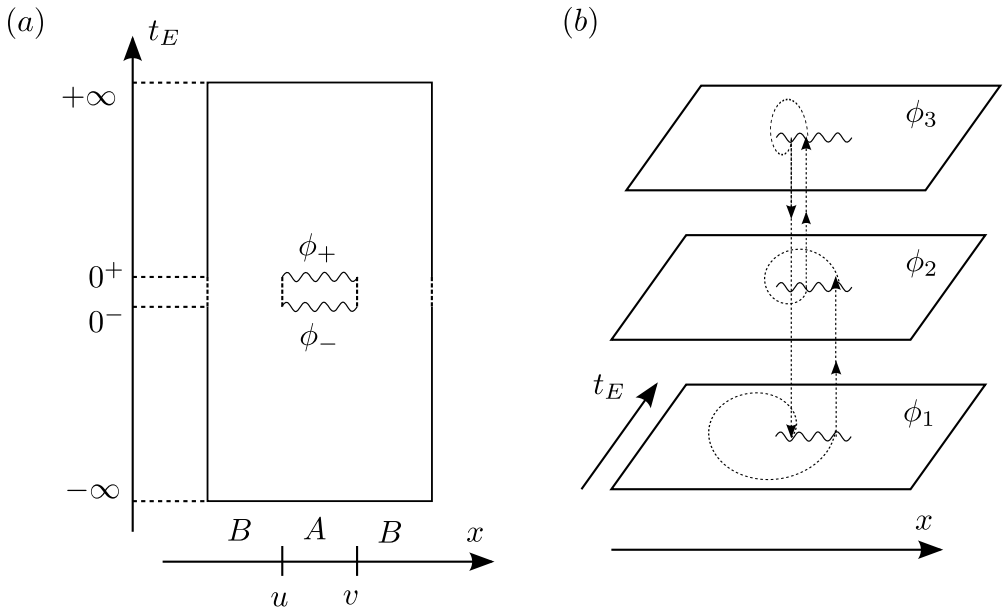}
\end{center}
\caption{
\label{fig: n-riemann.eps}
(a)
The path integral representation of the reduced
density matrix $[\rho_A]_{\phi_+\phi_-}$.
(b)
The $n$-sheeted Riemann surface $\mathcal{R}_n$.
(Here we take $n=3$ for simplicity.)
}
\end{figure}

This can be done in the path-integral formalism as follows.
First, assuming two-dimensional QFT just for simplicity,
we take
$A$ to be the single interval $x\in [u,v]$ at $t_E=0$ in
the flat Euclidean coordinates $(t_E,x)\in \mathbb{R}^{2}$. The
ground state wave functional $\Psi$ can be found by path-integrating
from $t_{E}=-\infty$ to $t_{E}=0$ in the Euclidean formalism \be
\Psi\left(\phi_0(x)\right)=
\int^{\phi(t_{E}=0,x)=\phi_0(x)}_{t_{E}=-\infty} D\phi~
e^{-S(\phi)} \ ,\ee where $\phi(t_E,x)$ denotes the field which defines
the 2D QFT.
The values of the field at the boundary $\phi_0$ depends
on the spatial coordinate $x$. The total density matrix $\rho$ is
given by two copies of the wave functional
$[\rho]_{\phi_0\phi_0'}=\Psi(\phi_0)\bar{\Psi}(\phi'_0)$. The
complex conjugate one $\bar{\Psi}$ can be obtained by
path-integrating from $t_{E}=\infty$ to $t_{E}=0$. To obtain the
reduced density matrix $\rho_A$, we need to integrate $\phi_0$ on
$B$ with the condition $\phi_{0}(x)=\phi'_0(x)$ when $x\in B$
\be
[\rho_A]_{\phi_+ \phi_-}=(Z_{1})^{-1}\int
^{t_E=\infty}_{t_{E}=-\infty} D\phi~ e^{-S(\phi)}\prod_{x\in
A}\delta\left(\phi(+0,x)-\phi_+(x)\right)\cdot
\delta\left(\phi(-0,x)-\phi_-(x)\right) \ , \label{pathrho}\ee where
$Z_1$ is the vacuum partition function on $\mathbb{R}^{2}$ and we
multiply its inverse in order to normalize $\rho_A$ such that
$\mathrm{tr}_A\,\rho_A=1$. This computation is sketched in Fig.\
\ref{fig: n-riemann.eps} (a).

To find $\mathrm{tr}_A\,\rho_A^n$, we can prepare $n$ copies of
(\ref{pathrho}) \be [\rho_A]_{\phi_{1+}
\phi_{1-}}[\rho_A]_{\phi_{2+} \phi_{2-}}\ddd [\rho_A]_{\phi_{n+}
\phi_{n-}} \ ,\ee and take the trace successively. In the path-integral
formalism this is realized by gluing $\{\phi_{i\pm}(x)\}$ as
$\phi_{i-}(x)=\phi_{(i+1)+}(x)$ ($i=1,2,\ddd,n$) and integrating
$\phi_{i+}(x)$. In this way, $\mathrm{tr}_A\,\rho_A^n$ is given in
terms of the path-integral on an $n$-sheeted Riemann surface
${\mathcal R}_n$ (see Fig.\ \ref{fig: n-riemann.eps} (b)) \be
\mathrm{tr}_A\,\rho_A^n=(Z_1)^{-n}\int_{(t_E,x)\in {\mathcal R}_n}
D\phi~ e^{-S(\phi)}\equiv \f{Z_n}{(Z_1)^n} \ . \label{nsheetp}  \ee

Though we have assumed two-dimensional
QFTs so far, it can be straightforwardly generalized to
higher dimensions. Then $Z_n$ becomes a partition function on a singular space which is
obtained by gluing $n$ copies of the original space along $\de A$. It has a negative deficit
angle $2\pi (1-n)$ along the surface $\de A$. This becomes two end points of the cut in the
two-dimensional example.

In two-dimensional CFTs, it is possible to analytically calculate
(\ref{nsheetp}) to find the formula (\ref{enttdcft})
\cite{HLW,Cardy,Casini05a,Casini05b,Casini:2007bt} as it essentially
becomes products of two point functions of twisted vertex operators.
In the case where the subsystem $A$ consists of multiple intervals,
recent discussions are available in
\cite{Casini:2008wt,Caraglio:2008pk,CCT}. However, in higher
dimensions, analytical calculations of $S_A$ become very
complicated. Below we list some recent progresses in this direction.
The analytical results when $A$ and $B$ divide a flat spacetime
along a flat plane has been found in
\cite{Kabat,Cardy,Nishioka:2006gr} (see also \cite{Fu} for the contributions 
from cusps). 
When the subsystem $A$ is a
straight strip, numerical results are available in
\cite{Casini05b,RuTaL}. Also, if we assume that $A$ includes a cusp
singularity, then the entanglement entropy of $(2+1)$-dimensional
QFT has a logarithmic term. This is evaluated in
\cite{Casini:2006hu,Casini:2008as}. Lattice calculations in
non-abelian gauge theories have also been performed in
\cite{Velytsky:2008rs, Buividovich:2008kq, Buividovich:2008gq,
Buividovich:2008yv}. The calculations at 2D quantum Lifshitz fixed points
have been performed in \cite{Frad}. The entanglement entropy in the $O(N)$ model
has been computed by employing the $\ep$ expansion in \cite{MFS},
quite recently. However, there have only been very few analytical
calculations of $S_A$ in generic interacting QFTs in dimensions
greater than two. Thus our holographic approach which can be applied
to strongly coupled theories will provide a powerful complementary
method.

\subsection{Entanglement Entropy and Black Holes}
\label{EEBH}
\hspace{5mm}
It may be interesting to notice that the area law (\ref{divarea})
looks very similar to the Bekenstein-Hawking entropy
of black holes
which is proportional to the area of the event horizon (\ref{BHF}).
Intuitively, we can regard $S_A$
as the entropy for an observer who is only accessible to the
subsystem $A$ and cannot receive any signals from $B$. In this
sense, the subsystem $B$ is analogous to the inside of a black hole
horizon for an observer sitting in $A$, i.e., outside of the
horizon. Indeed, this similarity was an original motivation of the
entanglement entropy \cite{Thooft,Bombelli,Srednicki}.

However, one may immediately notice the discrepancy between them.
The entanglement entropy is proportional to the number of matter fields
, while the black hole entropy is
not. Also the former includes ultraviolet
divergences as opposed to the latter.  The correct statement of this
relation turns out to be that quantum corrections to the
black hole entropy in the presence of matter fields is equal to
the entanglement entropy \cite{SuUg,FPST,Jacobson:1994iw,Solok}. In the setup
of the induced gravity, where Einstein-Hilbert action is assumed to be
all generated from the quantum corrections to matter fields, we can fully identify
the black hole entropy with the entanglement entropy \cite{Jacobson:1994iw}.
This relation has been reconsidered in \cite{Dvali:2008jb} recently by proposing the existence of
a new gravitational cut off.

In our holographic argument below, we will present an identification of the entanglement entropy
in $(d+1)$-dimensional QFT
with a certain geometrical quantity in $(d+2)$-dimensional gravity, which can be regarded as a generalization of
the black hole entropy. This relation holds whenever holography dual of the QFT exists.
In particular case of brane-world setup, our identification turns out to be reduced to the
mentioned equivalence between the black hole entropy and entanglement entropy
in the induced gravity \cite{HMS,Emparan:2006ni,Solodukhin:2006xv}
as we will review in Sec.\ \ref{secbh} and Sec.\ \ref{seccov}.

\section{Holographic Entanglement Entropy}
\label{holographic} \setcounter{equation}{0} \hspace{5mm}
Here we would like to explain the holographic calculation of the entanglement entropy.
In order to simplify the notations and reduce ambiguities,
we consider the setup of the AdS/CFT correspondence, though it will be rather straightforward to
extend our results to general holographic setups.

The AdS/CFT correspondence argues that (quantum) gravity in
the $(d+2)$-dimensional anti de-Sitter space AdS$_{d+2}$ is equivalent to a $(d+1)$-dimensional conformal
field theory CFT$_{d+1}$ \cite{Maldacena}. Below we mainly employ the
Poincare metric of AdS$_{d+2}$ with radius $R$:
\be ds^2=R^2~\f{dz^2-dx_0^2+
\sum_{i=1}^{d-1}dx_i^2}{z^2} \ .
\label{Poincare} \ee
The dual CFT$_{d+1}$ is supposed to live on the boundary of AdS$_{d+2}$
which is $R^{1,d}$
at $z\to 0$ spanned by the coordinates $(x^0,x^i)$. The extra coordinate
$z$ in AdS$_{d+2}$ is interpreted as the length scale of the dual CFT$_{d+1}$
in the RG sense.
Since the metric diverges
in the limit $z\to 0$, we put a cut off by imposing $z \geq a$. Then the boundary is situated
at $z=a$ and this cut off $a$ is identified with the ultraviolet cut off in the dual CFT.
Under this interpretation, a fundamental principle of AdS/CFT, known as the bulk to boundary
relation \cite{ADSGKP,ADSWitten}, is simply expressed by the equivalence of the partition
functions in both theories
\be
Z_{CFT}=Z_{AdS\ Gravity} \ .\label{btob}
\ee
Since (non-normalizable) perturbations in the AdS background by exciting fields in the AdS side
are dual to
the shift of background in the CFT side, we can compute the correlation functions in the CFT
by taking the derivatives with respect to the perturbations. In generic parameter regions,
the gravity should be treated in string theory to take the quantum corrections into account.
Nevertheless, in particular interesting limit, typically strong coupling limit of CFT,
the quantum corrections become negligible and we can employ supergravity to describe AdS
spaces. Moreover, most of our examples shown in this article are simple enough that we can apply
general relativity. In this situation, the right-hand side of (\ref{btob}) is reduced to the
exponential $e^{-S_{EH}}$ of the on-shell Einstein-Hilbert action.

So far we applied the AdS/CFT to the pure AdS spacetime (\ref{Poincare}).
However, the AdS/CFT can be applied to any asymptotically AdS spacetimes
including the AdS black holes.

\subsection{Holographic Formula}
\label{general proposal} \hspace{5mm}
Now we are in a position to present how to calculate the entanglement
entropy in CFT$_{d+1}$ from the gravity on AdS$_{d+2}$. This argument here
can be straightforwardly generalized to any static backgrounds.

\begin{figure}
\begin{center}
\includegraphics[width=8cm,clip]{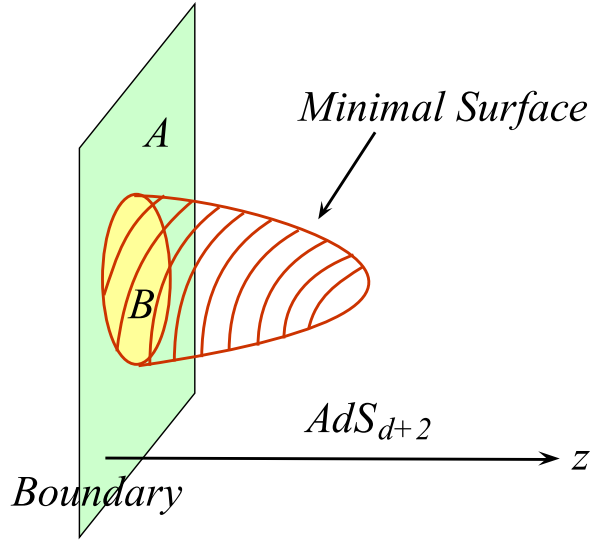}
\end{center}
\caption{
\label{Fig:holo}
The holographic calculation of entanglement entropy via AdS/CFT.}
\end{figure}

To define the entanglement entropy  in the CFT$_{d+1}$,
we divide the (boundary) time slice $N$
into $A$ and $B$ as we explained before (see Fig.\ \ref{Fig:holo}).
In the Poincare coordinate (\ref{Poincare}), we are setting $N=R^d$ and
the CFT$_{d+1}$ is supposed to live on the boundary $z=a\to 0$
of AdS$_{d+2}$.
To have its dual gravity picture, we need to extend
this division $N=A\cup B$ to the time slice $M$ of  the
bulk spacetime. In the setup (\ref{Poincare}),
$M$ is the $(d+1)$-dimensional hyperbolic spacetime $H_{d+1}$.
Thus we extend $\partial A$ to a surface $\gamma_A$ in the entire $M$ such
that $\partial \gamma_A=\partial A$. Notice that this is a surface
in the time slice $M$, which is a Euclidean manifold. Of course,
there are infinitely many different choices of $\gamma_A$. We claim
that we have to choose the minimal area surface among them. This means
that we require that the variation of the area functional vanishes;
if there are multiple solutions, we choose the one whose area takes
the minimum value. This procedure singles out a unique minimal surface
and we call this $\gamma_A$ again
(see Fig.\ \ref{Fig:holo}).

In this setup we propose that
the entanglement entropy $S_A$ in CFT$_{d+1}$ can be computed from
the following formula \cite{RuTa,RuTaL}
\begin{equation} S_{A}=\frac{{\rm Area}(\gamma_{A})}{4G^{(d+2)}_N} \ .
\label{arealaw}
\end{equation}
We stress again that the manifold $\gamma_{A}$ is
the $d$-dimensional minimal area
surface in AdS$_{d+2}$ whose boundary is given by $\de A$. Its area
is denoted by ${\rm Area}(\gamma_{A})$. Also $G^{(d+2)}_{N}$ is the
$(d+2)$-dimensional Newton constant of the AdS gravity.
We can easily show that the leading
divergence $\sim a^{-(d-1)}$ in (\ref{arealaw}) is proportional to
the area of the boundary $\de A$ and this immediately reproduces the
area law property (\ref{divarea}).

The appearance of the formula (\ref{arealaw}) looks very similar to the area law
of the Bekenstein-Hawking formula (\ref{BHF}) of black hole entropy.
Indeed, we can regard our formula (\ref{arealaw})
as a generalization of (\ref{BHF}) because in the presence of
event horizon
such as the AdS Schwarzschild black hole solutions, the minimal surface tends
to wrap the horizon. Refer to Sec.\ \ref{AdS3/CFT2BH} for more details.

This formula (\ref{arealaw}) was originally motivated by the
following intuitive interpretation\cite{RuTa,RuTaL}.
Since the entanglement entropy $S_A$ is defined by smearing out the
region $B$, the entropy is considered to be the one for an observer
in $A$ who is not accessible to $B$. The smearing process produces
the fuzziness for the observer and that should be measured
 by $S_A$. In the higher-dimensional
perspective of the AdS space, the fuzziness appears by hiding a
part of the bulk space AdS$_{d+2}$ inside an imaginary horizon,
which we call $\gamma$.  It is clear that $\gamma$ covers the
smeared region $B$ from the inside of the AdS space and thus we find
$\de \gamma=\de B(=\de A)$. To make this imaginary horizon more
precise, we can employ the argument of the entropy bound \cite{Bousso}.
This idea, roughly speaking, claims that the entropy contained in a certain space
is bounded by the area of its surface (for details, see Sec.\ \ref{seccov}).
 To choose the minimal
surface as in (\ref{arealaw}) means that we are seeking the severest
entropy bound \cite{holography,BiSu,Bousso} so that it has a chance to
saturate the bound. Refer to Sec.\ \ref{seccov} for more details.

In the above we implicitly assume that the subsystem $A$ is a connected
manifold. When it is disconnected,
we need to extend the holographic formula properly.
A candidate of formula in disconnected cases has
been proposed in \cite{Hubeny:2007re} based on the strong subadditivity.

\subsection{Heuristic Derivation of Holographic Formula}
\hspace{5mm}
In principle, we should be able to perform the holographic calculation of
the entanglement entropy
based on the first principle of the AdS/CFT correspondence known
as the bulk to boundary relation (\ref{btob}). In the CFT side, the entanglement entropy
can be found if we can compute the partition function on the $(d+1)$-dimensional
$n$-sheeted space (\ref{nsheetp}) via
the formula (\ref{deri}). This space, called ${\cal R}_n$,
is characterized by the presence of the deficit angle
$\delta=2\pi(1-n)$ on the surface $\de A$.
Therefore we need to find a $(d+2)$-dimensional
back reacted geometry ${\cal S}_n$ by solving the Einstein equation
with the negative cosmological constant such that its metric approaches
to that of ${\cal R}_n$ at the boundary $z\to 0$. This is a technically complicated mathematical
problem if we try to solve it directly and has
not been completely solved at present.

To circumvent this situation, we make a following natural assumption following \cite{Fursaev:2006ih}:
the back reacted geometry ${\cal S}_n$ is given by a $n$-sheeted AdS$_{d+2}$, which is defined by
putting the deficit angle $\delta$ localized on a codimension two surface $\gamma_A$.
This is clearly true in the three-dimensional pure gravity as the solution to the Einstein equation should
be locally the same as AdS$_3$. However, this is not trivially obvious in higher dimensions.
Under this assumption,
the Ricci scalar behaves like a delta
function
\begin{equation}
R=4\pi (1-n)\delta(\gamma_A)+R^{(0)} \ ,
\end{equation}
where
$\delta(\gamma_A)$ is the delta function localized on $\gamma_A$,
$\delta(\gamma_A)=\infty$ for $x\in \gamma_A$
whereas $\delta(\gamma_A)=0$ otherwise,
and $R^{(0)}$ is that of the pure AdS$_{d+2}$. Then we plug this in the supergravity action
\begin{equation}
S_{AdS}=-\frac{1}{16\pi G_N^{(d+2)}}\int_M
dx^{d+2}\sqrt{g}(R+\Lambda)+\cdot\cdot\cdot \ ,
\end{equation}
where we only make explicit the bulk Einstein-Hilbert action. This is because the
other parts omitted in the above such as kinetic terms of scalars, lead to extensive
terms which are proportional to $n$ and are canceled in the ratio (\ref{nsheetp}).
Now the bulk to boundary relation (\ref{btob}) equates the partition function of CFT with the one of
AdS gravity. Thus we can holographically calculate the entanglement
entropy $S_A$ as follows
\begin{equation}\label{HolEE}
S_A=-\frac{\partial }{\partial n}\log
\mbox{Tr}\rho_A^n|_{n=1}=-\frac{\partial }{\partial
n}\left[\frac{(1-n)\mbox{Area}(\gamma_A)}{4G_N^{d+2}}\right]_{n=1}
=\frac{\mbox{Area}(\gamma_A)}{4G_N^{d+2}} \ .
\end{equation}
The action principle in the gravity theory requires that $\gamma_A$ is
the minimal area surface. In this way, we reproduced
our holographic formula (\ref{arealaw}) \cite{Fursaev:2006ih}. Notice that the presence of
non-trivial minimal surfaces is an well-established property of asymptotically
AdS spaces.

In this derivation of the holographic formula, the assumption about the back reacted geometry
${\cal S}_n$ has been crucial. This assumption is clearly satisfied for the three-dimensional pure gravity
as we noticed in the above (see also \cite{Michalogiorgakis:2008kk} for detailed analysis).
To explore this issue in higher-dimensional AdS/CFT, hopefully demonstrating the proof of (\ref{arealaw}),
is one of the most important future problems in
the holographic entanglement entropy\footnote{
Recently, a subtle disagreement
about the logarithmic term of the entanglement entropy
between the holographic result (\ref{arealaw})
and the CFT result is pointed out
based on the anomaly analysis
in \cite{Schwimmer:2008yh}. This occurs
when the extrinsic curvature of $\de A$
is non-vanishing, where the geometric analysis gets quite complicated.
However, it is possible that this problem
arises from subtle differential geometric calculations
in the presence of deficit angles as suggested in
\cite{Solodukhin:2008dh}, where
agreements between the gravity and CFT sides have been
observed for the logarithmic term.
This issue should clearly deserve detailed future analysis.}.

\subsection{Holographic Proof of Strong Subadditivity}
\label{stsub}
\hspace{5mm}
One of the most important properties of the entanglement entropy is the strong subadditivity \cite{LiRu}
given by the inequalities (\ref{Stronga}) and (\ref{Strongb}).
This represents the concavity of the entropy and is somehow analogous to the second law of
thermodynamics. Actually, it is possible to check that our holographic formula (\ref{arealaw})
satisfies this property in a rather simple argument as shown in \cite{Headrick:2007km}
(see also \cite{Hirata:2006jx} for the explicit numerical studies).

\begin{figure}
\begin{center}
\includegraphics[width=8cm,clip]{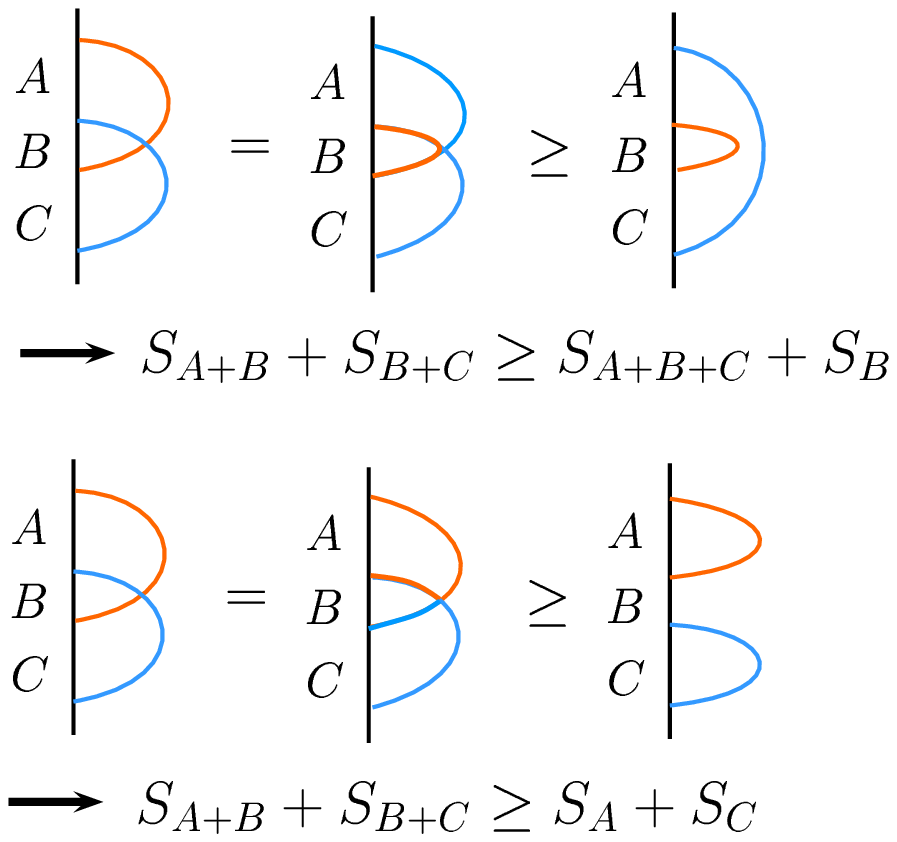}
\end{center}
\caption{
\label{Fig:SSA}
A holographic proof of the strong subadditivity of the entanglement entropy. To make the figures simple,
we project the time slice of a $(d+2)$-dimensional AdS space onto a two-dimensional plane. This
simplification does not change our result.}
\end{figure}

Let us start with three regions $A$, $B$ and $C$ on a time slice of a given CFT so that there are no
overlaps between them. We extend this boundary setup toward the bulk AdS
(see Fig.\ \ref{Fig:SSA}).
Consider the entanglement entropy $S_{A+B}$ and $S_{B+C}$. In the holographic
description (\ref{arealaw}), they are given by the areas of minimal area surfaces $\gamma_{A+B}$ and
$\gamma_{B+C}$ which satisfy $\de \gamma_{A+B}=\de (A+B)$ and $\de \gamma_{B+C}=\de (B+C)$ as before.
Then it is easy to see that we can divide these two minimal surfaces into four pieces and
recombine into (i) two surfaces $\gamma'_B$ and $\gamma'_{A+B+C}$
or (ii) two surfaces $\gamma'_A$ and $\gamma'_C$, corresponding to two different ways of the
recombination. Here we again
meant $\gamma'_X$ is a surface which satisfies $\de \gamma'_X=\de X$.
Since in general $\gamma'_X$s are not minimal area surface, we have Area$(\gamma'_X)\geq $Area$(\gamma_X)$.
Therefore, as we can easily find from Fig.\ \ref{Fig:SSA}, this argument immediately leads to
\ba
&& \!\!\!\! \mbox{Area}(\gamma_{A+B})+\mbox{Area}(\gamma_{B+C})
=\mbox{Area}(\gamma'_{B})+\mbox{Area}(\gamma'_{A+B+C})\geq
\mbox{Area}(\gamma_{B})+\mbox{Area}(\gamma_{A+B+C}) \ , \no
&& \!\!\!\! \mbox{Area}(\gamma_{A+B})+\mbox{Area}(\gamma_{B+C})
=\mbox{Area}(\gamma'_{A})+\mbox{Area}(\gamma'_{C})\geq
\mbox{Area}(\gamma_{A})+\mbox{Area}(\gamma_{C}) \ .
\ea
In this way, we are able to check the strong subadditivity (\ref{Stronga}) and (\ref{Strongb}).
Analogous inequalities have been discussed in \cite{Hirata:2008ms} for the holographic
Wilson loops.

\subsection{Entanglement Entropy from AdS$_3$/CFT$_2$}
\label{AdS3/CFT2} \hspace{5mm}
Consider AdS$_3/$CFT$_2$ as one of the simplest setups of AdS/CFT.
Since the entanglement entropy in two-dimensional CFT can be analytically obtained
as we mentioned, we can test our holographic formula explicitly.
The central charge of CFT is related to radius of AdS$_3$ \cite{BH} via
\be
c=\f{3R}{2G^{(3)}_N} \ .\label{centralads}\ee

\subsubsection{Entanglement Entropy in CFT$_2$ at Zero Temperature}
\hspace{5mm}
We are interested in the entanglement entropy $S_A$ in an infinitely long system when
$A$ is an interval of length $l$. To compute this via the holographic formula (\ref{arealaw}),
we need to find a geodesics between the two points $(x^1,z)=(-l/2,a)$ and $(x^1,z)=(l/2,a)$
in the Poincare coordinate (\ref{Poincare}). It is actually given by the half circle
  \be (x,z)=\f{l}{2}(\cos s,\sin s) \ ,\ \ \
\ (\ep\leq s \leq \pi-\ep) \ , \label{geodesicpp}\ee
where $\ep=\f{2a}{l}$.
The length of $\gamma_{A}$
can be found as \be {\rm Length}(\gamma_A)=2R\int^{\pi/2}_\ep \f{ds}{\sin
s}=-2R\log(\ep/2)=2R\log\f{l}{a} \ .\label{lenghthpo} \ee Finally the
entropy can be obtained as follows \be S_A=\f{{\rm
Length}(\gamma_A)}{4G^{(3)}_N}=\f{c}{3}\log\f{l}{a} \ .\label{entrppp}\ee
This perfectly agrees with the result (\ref{enttdcft}) in the CFT side \cite{RuTa,RuTaL}.
By starting from the global coordinate of AdS$_3$ we can also derive the result
(\ref{entropyone}) similarly.

\subsubsection{Entanglement Entropy in CFT$_2$ at Finite Temperature}
\label{AdS3/CFT2BH}
\hspace{5mm} Next we consider how to explain the entanglement
entropy (\ref{entropytemp}) at finite temperature $T=\beta^{-1}$
from the viewpoint of the AdS/CFT correspondence.
We assume that
the spatial length of the total system $L$ is infinite i.e.
$\beta/L\ll 1$. In such a high temperature region, the gravity
dual of the conformal field theory is described by the Euclidean BTZ
black hole \cite{BTZ}. Its metric looks like \be
ds^2=(r^2-r_+^2)d\tau^2+ \f{R^2}{r^2-r^2_{+}}dr^2+r^2 d\vp^2 \ .
\label{btzmet}\ee The Euclidean time is compactified as
$\tau\sim\tau+\f{2\pi R}{r_+}$ to obtain a smooth geometry. We also
impose the periodicity $\vp\sim \vp+2\pi$. By taking the boundary
limit $r\to \infty$, we find the relation between the boundary CFT
and the geometry (\ref{btzmet}) \be \f{\beta}{L}=\f{R}{r_{+}}\ll 1 \ .
\label{relationbtz}\ee

The subsystem for which we consider the entanglement entropy is
given by $0\leq \vp\leq 2\pi l/L$ at the boundary. Then by extending
our formula (\ref{arealaw}) to asymptotically AdS spaces,
 the entropy can be computed from the
length of the space-like geodesic starting from $\vp=0$ and ending
at $\vp=2\pi l/L$ at the boundary $r=r_0\to \infty$ at a fixed time.
This geodesic distance can be found analytically as
\be \cosh\left(\f{\mbox{Length}(\gamma_A)}{R}\right)
=1+\f{2r_0^2}{r_+^2}\sinh^2\left(\f{\pi
l}{\beta}\right) \ .\label{delltwo} \ee
The relation between the cut off $a$ in CFT and
the one $r_0$ of AdS is given by $\f{r_0}{r_+}=\f{\beta}{a}$.
Then it is easy to see that our area law
(\ref{arealaw}) precisely reproduces the known CFT result
(\ref{entropytemp}).

\begin{figure}
\begin{center}
\includegraphics[height=4.4cm,clip]{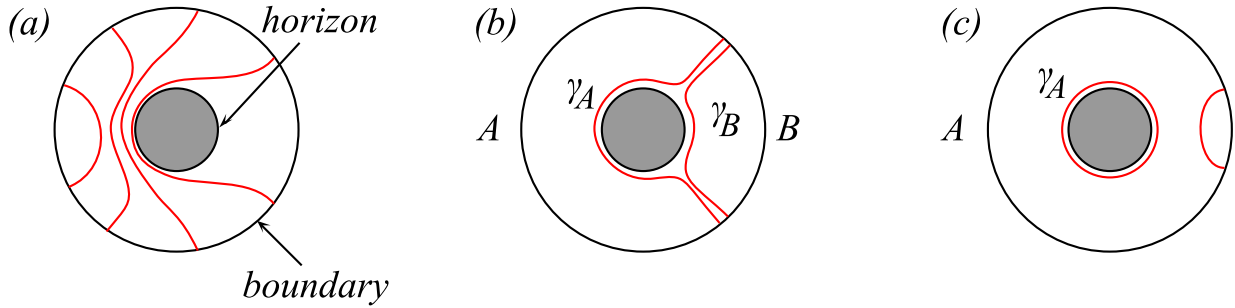}
\end{center}
\caption{ \label{fig: ads_blackhole.eps} (a) Minimal surfaces
$\gamma_A$ in the BTZ black hole for various sizes of $A$. (b)
$\gamma_A$ and $\gamma_B$ wrap the different parts of the horizon.
(c) When $\de A$ gets larger, $\gamma_A$ is separated into
two parts: one is wrapped on the horizon and the other localized near the boundary.}
\end{figure}

It is also useful to understand these calculations geometrically.
The geodesic line in the BTZ black hole takes the form shown in
Fig.\ \ref{fig: ads_blackhole.eps}(a). When the size of $A$ is
small, it is almost the same as the one in the ordinary AdS$_3$. As
the size becomes large, the turning point approaches the horizon and
eventually, the geodesic line covers a part of the horizon. This is
the reason why we find a thermal extensive behavior of the entropy when
$l/\beta\gg 1$ in (\ref{entropytemp}). The thermal entropy in a
conformal field theory is dual to the black hole entropy in its
gravity description via the AdS/CFT correspondence. In the presence
of a horizon, it is clear that $S_A$ is not equal to $S_B$ (remember
$B$ is the complement of $A$) since the corresponding geodesic lines
wrap different parts of the horizon (see Fig.\ \ref{fig:
ads_blackhole.eps}(b)).
This is a typical property of
the entanglement entropy at finite temperature
as we mentioned in Sec.\ \ref{basics}.
We also expect that when $A$ becomes very large before it coincides
with the total system, $\gamma_A$ becomes separated into the horizon circle and
a small half circle localized on the boundary  (see Fig.\ \ref{fig:
ads_blackhole.eps}(c)). We can indeed confirm that this indeed happens in
the dual CFT result (\ref{toth}) as shown in \cite{Azeyanagi:2007bj}.

\subsubsection{Massive Deformation}
\label{massive deform} \hspace{5mm}
Massive quantum field theories can
be obtained by perturbing two-dimensional conformal field theories by
relevant perturbations. In the dual gravity side, this corresponds to
an IR deformation of AdS$_3$ space. As in the
well-known examples \cite{Witten,KlSt,PoSt,MaNu} of confining gauge theories,
we expect the massive deformation caps off the IR region $z>z_{IR}$.

Consider an $(1+1)$-dimensional infinite system divided into two
semi-infinite pieces and define the subsystem $A$ to be one of them.
The important quantity in the massive theory is the
correlation length $\xi$, which is identified with $\xi\sim z_{IR}$
in AdS/CFT. Since we assumed that the subsystem $A$ is
infinite, we should take a geodesic (\ref{geodesicpp}) with a large
value of $l(\gg\xi)$. Then the geodesic starts from the UV cutoff $z=a$
and ends at the IR cutoff $z=\xi$. Thus we can estimate the length of this
geodesic and finally the entanglement entropy as follows
 \be S_A=\f{{\rm
Length}(\gamma_A)}{4G^{(3)}_N}=
\f{R}{4G^{(3)}_N}\int^{2\xi/l}_{\ep=2a/l}\f{ds}{\sin
s}=\f{c}{6}\log\f{\xi}{a} \ .\label{entrpmas}
\ee This agrees with the known result \cite{Vidal03,Cardy} in
the $(1+1)$-dimensional quantum field theory.

\subsection{Holographic Entanglement Entropy in Higher Dimensions}
\label{AdS d+2/CFT d+1} 
\begin{figure}
\begin{center}
\includegraphics[width=15cm,clip]{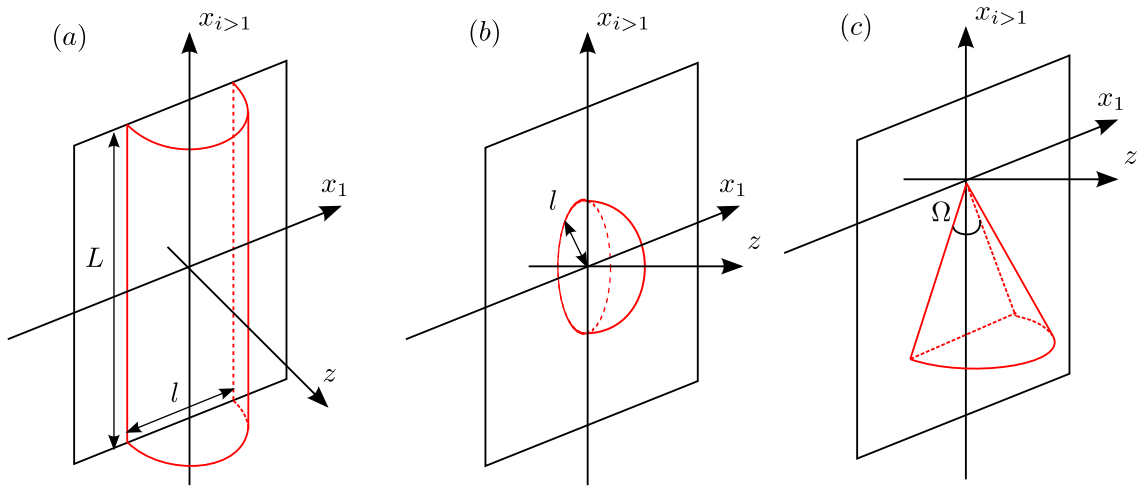}
\end{center}
\caption{
\label{fig: min_surf}
Minimal surfaces in
AdS$_{d+2}$: (a) $A_S$ (an infinite strip),
(b) $A_D$ (a disk)
and (c)  $A_W$ (a wedge).
}
\end{figure}

\hspace{5mm}
Next we turn to the holographic computation of the entanglement entropy in
higher dimensions. To obtain analytical results we assume that the subsystem
$A$ is given either by
(a) $d$-dimensional infinite strip (called $A_S$) with the width $l$
in one direction and the width $L(\to\infty)$ in other $d-1$ directions;
(b) $d$-dimensional disk (called $A_D$) with radius $l$; or
(c) $d$-dimensional wedge cone with a cusp with the angle $\Omega$ (called $A_W$).
We can find their corresponding minimal surfaces in AdS, explicitly as
depicted in Fig.\ \ref{fig: min_surf}.

\subsubsection{Entanglement Entropy for Infinite Strip $A_S$}
\hspace{5mm}
The holographic entanglement entropy (\ref{arealaw}) for $A_S$ is obtained
from (\ref{arealaw}) as follows \cite{RuTaL}
\be
S_{A_{S}}=\f{1}{4G^{(d+2)}_{N}}\left[\f{2R^d}{d-1}\left(\f{L}{a}\right)^{d-1}
-\f{2^d \pi^{d/2} R^{d}}{d-1}\left(\f{\Gamma(\f{d+1}{2d})}
{\Gamma(\f{1}{2d})}\right)^{d} \left(\f{L}{l}\right)^{ d-1}\right] \ ,
\label{areaone} \ee

Notice that the first divergent term is proportional to the area of
$\de A$ i.e. $L^{d-1}$ as we expect from the known area law in the
field theory computations (\ref{divarea}). The second term is finite
and thus is universal (i.e. does not depend on the cutoff). This is
the quantity which we can directly compare with the field theory
counterpart. The presence of these two terms agree with the field theoretic
results in \cite{Casini05b,RuTaL}. Notice that our result (\ref{areaone})
does not include subleading divergent terms ${\mathcal{O}}(a^{-d+3})$.

If we apply the above result (\ref{areaone}) to AdS$_5$/CFT$_4$,
we obtain the
following prediction of
the entanglement entropy
for the
${\mathcal N}=4$
$SU(N)$ super Yang-Mills theory
\ba S_{A_S}&=&\f{N^2L^2}{2\pi a^2}-2\s{\pi}
\left(\f{\Gamma\left(\f23\right)}{\Gamma\left(\f16\right)}
\right)^3\f{N^2L^2}{l^2} \ . \label{entrodthe} \ea Notice that this
is proportional to $N^2$ as expected since the number of fields in
the $SU(N)$ gauge theory is proportional to $N^2$. Moreover, for general even-dimensional
CFTs, we can show that
the holographic entanglement entropy for any choice of $A$ is always proportional to
the central charge \cite{RuTaL,Nishioka:2007zz}.

As we mentioned, it is intriguing to compare
the second finite term in (\ref{entrodthe}) to that obtained from field theoretic
calculations. The finite term in (\ref{entrodthe})
is numerically expressed as \be
S^{Sugra}_{A_S}|_{finite}\simeq-0.0510\cdot
\f{N^2L^2}{l^2} \ .\label{numeentr}\ee On the other hand, the free
field theory results can be obtained by employing the method first considered in
\cite{Casini05b}. The ${\mathcal N}=4$
super Yang-Mills consists of a gauge field $A_{\mu}$, six real
scalar fields $(\phi^1,\phi^2,\ddd,\phi^6)$ and four Majorana
fermions $(\psi^1_\ap,\psi^2_\ap,\psi^3_\ap,\psi^4_\ap)$. The contribution
from the gauge field is the same as those from two real scalar
fields \cite{Kabat}. In this way the total entropy in the free Yang-Mills theory
is the same as those from 8 real scalars and 4 Majorana fermions .
In this way, we eventually obtain the numerical estimation \cite{RuTa,RuTaL} \be S^{Free
YM}_{A_S}|_{finite}\simeq -(8\times 0.0049+4\times 0.0097)\cdot
\f{N^2L^2}{l^2}=-0.078 \cdot \f{N^2L^2}{l^2} \ .\label{numeentrr}\ee

We
observe that the free field result (\ref{numeentrr}) is larger
than the one (\ref{numeentr}) in the
gravity dual by roughly $50\%$. The deviation itself is anticipated
since our holographic computation should give the result in the strongly coupling
limit and the entanglement entropy is not a protected quantity which does not depend on the
coupling constant. This
situation is very similar to the computation of thermal entropy
\cite{GKP}, where we have a similar discrepancy (so-called
$\f{4}{3}$ problem). The fact that the discrepancy is of order one
also in our computation can be thought as an encouraging evidence
for our argument.

We may also apply this holographic calculation to gauge theories in different dimensions.
We can find holographic results of the entanglement entropy in
\cite{Arean:2008az} for 2D $\CN =(4,4)$ Yang-Mills,
in  \cite{Ramallo:2008ew} for 3D $\CN =4$ Yang-Mills, and in \cite{Nishioka:2008gz}
for 3D $\CN =6$ Chern-Simons theory. Refer to\cite{Barbon:2008sr,Barbon:2008ut}
for the analysis in the presence of gauge fluxes.

\subsubsection{Entanglement Entropy for Circular Disk $A_D$}
\label{EE for A_D}
\hspace{5mm}
The holographic entanglement entropy (\ref{arealaw}) for $A_D$ is found as follows \cite{RuTaL}
\begin{eqnarray}
S_{A_{D}} &=&
\f{2\pi^{d/2}R^d}{4G^{(d+2)}_{N}\Gamma(d/2)}
\int^1_{a/l}dy \f{(1-y^2)^{(d-2)/2}}{y^d}  \nonumber \\
&=&
  p_1  \left(l/a\right)^{d-1}
+ p_3 \left(l/a \right)^{d-3}
+\cdots  \label{areatwo}  \\
&&
\cdots +\left\{
\begin{array}{ll}
\displaystyle p_{d-1}\left(l/a\right) + p_d + \mathcal{O}(a/l) \ , &
 \mbox{$d$: even} \ ,   \\
\displaystyle p_{d-2} \left(l/a\right)^{2} + q \log
\left(l/a\right)+ \mathcal{O}(1) \ ,
&  \mbox{$d$: odd} \ ,   \\
\end{array}
\right.
 \nonumber
\end{eqnarray}
where the coefficients are defined by \ba p_1/C &=& (d-1)^{-1} \ ,\ \
p_3/C = - (d-2)/[2(d-3)],\ \ \ddd \no p_d/C &=&
(2\s{\pi})^{-1}\Gamma(d/2) \Gamma\left((1-d)/2\right) \ \
(\mbox{if}\ \  d=\mbox{even}) \ ,\no q/C &=&
(-)^{(d-1)/2}(d-2)!!/(d-1)!! \ \  (\mbox{if}\ \  d=\mbox{odd}) \ ,
\no
&&\mbox{where}\ \ \  C\equiv
\f{\pi^{d/2}R^d}{2G^{d+2}_{N}\Gamma(d/2)} \ .  \label{entropydg}\ea

We notice that the result (\ref{entropydg}) includes a leading UV
divergent term $\sim a^{-d+1}$ and its coefficient
 is proportional to the area of the boundary $\de A$ as expected from
the area law \cite{Bombelli,Srednicki} in the field theories
(\ref{divarea}). We have also subleading divergent terms which
reflects the form of the boundary $\de A$.

In particular, we prefer a physical quantity that is independent of
the cutoff (i.e.~universal). The final term in (\ref{entropydg}) has
such a property. When $d$ is even, it is given by a constant $p_d$.
This seems to be somewhat analogous to the topological entanglement
entropy (or quantum dimension) in $(2+1)$ D
topological field theories \cite{Kitaev05, Levin05}, though our
theory is not topological.
On the other hand, when $d$ is odd, the coefficient $q$ of the
logarithmic term $\sim \log(l/a)$ is universal as
in the 2D
case (\ref{enttdcft}). In higher-dimensional CFTs, we can show that
$q$ is proportional to a certain linear combination of central charges.

This result is based on an explicit calculation when $A=A_D$.
However, from
\cite{GrWi},
we find that the behavior
(\ref{areatwo}) is also true for any compact submanifold $A$ with
different coefficient $p_k$ and $q$ depending on the shape of $A$.

\subsubsection{Entanglement Entropy and Cusps}
\label{EE for cusps}
In the third example $A=A_W$ with $d=2$, we can obtain the following result \cite{Hirata:2006jx}
\be
S_A=\f{R^2}{4G^{(4)}_N}
\left(\f{2L}{a}-2f(\Omega)\log\f{L}{a}\right)\ ,\label{cuspee}
\ee
where the function $f(\Omega)$ is given by
\be
f(\Omega)=\int^{\infty}_0
dz\left[1-\s{\f{z^2+g_0^2+1}{z^2+2g_0^2+1}}\right]\ ,\label{funca}
\ee
with
\be
\f{\Omega}{2}=g_0\s{1+g_0^2}\int^\infty_0\f{dz}{(z^2+g_0^2)\s{(z^2+g_0^2+1)(z^2+2g_0^2+1)}}\
.
\ee
The presence of the characteristic logarithmic term is due to the presence of the cusp singularity
of the space $\de A$. We can show that
$f(\Omega)$ is a convex (or $-f(\Omega)$ is concave) function. This property $f''(\Omega)\geq 0$
is actually what the strong subadditivity requires (for details refer to \cite{Hirata:2006jx}).

In free scalar and fermion theories, $S_A$ with $A=A_W$ has been computed in $2+1$ dimensions
and the same scaling structure (\ref{cuspee}) has been found \cite{Casini:2006hu,Casini:2008as}, where
the form of the function $f(\Omega)$ also turns out to agree semi-quantitatively with our strong coupling
limit prediction (\ref{funca}).

\section{Entanglement Entropy as an Order Parameter}
\setcounter{equation}{0}
\label{secpha}
\hspace{5mm}

In recent discussions in condensed matter physics,
the entanglement entropy is expected to play
a role of an appropriate order parameter
describing quantum phases and phase transitions.
For example,
it can be particularly useful for
a system which realizes a topological order,
such as fractional quantum Hall systems.
At low energies,
such systems can be described by a topological field theory,
and the correlation functions are not useful order parameter
as they are trivial.
However, the entanglement entropy can capture
important information of the topological
ground state \cite{Levin05,Kitaev05}.
Also it is interesting to note that the entanglement entropy
has been employed to estimate
efficiency of a numerical algorithm, such as DMRG \cite{DMRG},
which makes use of the (reduced) density matrix
as a criterion to discard unimportant information.

This is because the entanglement entropy can measure the amount of lost information
by the coarse graining procedure or equally the renormalization flow \cite{Vidal03,Gaite,Orus05}.

The main purpose of this section is to apply the entanglement
entropy to the confinement/deconfinement transition of gauge theories
\cite{Nishioka:2006gr,Klebanov:2007ws,Faraggi:2007fu,Bah:2008cj}.
It is much easier to employ our holographic calculation as we need to
deal with strongly
coupled gauge theories.

\subsection{Confinement/Deconfinement Transition}\hspace{5mm}
One of the most interesting applications of the entanglement entropy is that
it can be used as an order parameter for the confinement/deconfinement phase
transition in the confining gauge theory. When we divide one of the spatial direction
into a line segment with length $l$ and its complement, the entanglement entropy
between the two regions measures the effective degrees of freedom at the energy scale
$\Lambda \sim 1/l$. Then, in the confining gauge theory, the behavior entanglement entropy
should become trivial (i.e. $S_A$ approaches to a constant) as $l$ becomes large, i.e. the infrared
limit $\Lambda\to 0$.

Such a transition can be captured by the holographic entanglement entropy if there
are confining backgrounds dual to the confining gauge theories
\cite{Nishioka:2006gr,Klebanov:2007ws}. We find that there are two candidates for
the minimal surface with the same endpoints at the boundary in the confining background.
One is the two disconnected straight lines extending from the endpoints of the line segment
to inside the bulk, and the other is the curved line connecting these two points.
The connected curve correspond to the deconfinement phase in dual gauge theory because
the entanglement entropy depends on the length $l$, while
the disconnected lines independent of the length $l$ correspond to the confinement
phase.
In general, there is a critical length $l_c$ above which the disconnected lines are favored,
while below which the connected curve is favored, as we will see below explicitly.

For example, we consider the AdS soliton solution \cite{Witten}
\begin{align}
    ds^2 &= R^2\f{dr^2}{r^2 f(r)} + \f{r^2}{R^2} (-dt^2 + f(r) d\chi^2 + dx_1^2 + dx_2^2) \ ,
\end{align}
where $f(r) = 1 - r_0^4/r^4$ and the $\chi$ direction is compactified with the radius $L = \pi R^2/r_0$ to avoid
the conical singularity at $r=r_0$. This can be obtained from the double Wick rotation of the
AdS Schwarzschild solution. The dual gauge theory is $\CN =4$ super Yang-Mills on $R^{1,2}\times
S^1$, but the supersymmetry is broken due to the anti-periodic boundary condition for
fermions along the $\chi$ direction. Then the scalar fields acquire non-zero masses from
radiative corrections, and the theory becomes almost the same as the $(2+1)$-dimensional pure Yang-Mills,
which shows the confinement behavior \cite{Witten}.

To define the entanglement entropy, let us divide the boundary region into two parts $A$ and $B$:
$A$ is defined by $-l/2\le x_1 \le l/2, \ 0\le x_2 \le V(\to \infty)$ and $0\le \chi \le L$, and $B$ is the
complement of $A$.
The minimal surface $\g_A$, whose boundary coincides with the endpoint
$\p A$, can be obtained by minimizing
the area
\begin{align}\label{areaA}
    \text{Area} = LV \int_{-l/2}^{l/2}dx_1 \f{r}{R}\s{\left( \f{dr}{dx_1}\right)^2 + \f{r^4f(r)}{R^4}} \ .
\end{align}
Regarding $x_1$ as a time, then the energy conservation leads to
\begin{align}
    \f{dr}{dx_1} = \f{r^2}{R^2}\s{f(r)\left( \f{r^6f(r)}{r_*^6f(r_*)} - 1 \right)} \ ,
\end{align}
where $r_*$ is the minimal value of $r$.
When integrating this relation, we should also take the
boundary condition into account
\begin{align}\label{rtol}
    \f{l}{2} = \int_{r_*}^{r_\infty} dr \f{R^2}{r^2\s{f(r)\left( \f{r^6f(r)}{r_*^6f(r_*)} - 1 \right)}} \ ,
\end{align}
which relates $r_*$ with $l$. Here we introduced the UV cutoff at $r=r_\infty$.
After eliminating $l$ in (\ref{areaA}) and (\ref{rtol}), we find
the entanglement entropy as
\begin{align}
    S_A^{(con)} = \f{LV}{2RG_N^{(5)}} \int_{r_*}^{r_\infty} \f{r^4\s{f(r)}}{\s{r^6f(r) - r_*^6 f(r_*)}}
    \ .
\end{align}
It is important that $l$ is bounded from above
due to the relation
(\ref{rtol})
\begin{align}
    l \le l_{max} \simeq 0.22L \ .
\end{align}
Then, when $l$ becomes large, there is no minimal surface that connects the two boundaries of $\p A$.
Instead, the disconnected straight lines actually dominate before $l$ becomes greater than $l_{max}$.
The entanglement entropy is easy to be found
\begin{align}
    S_A^{(discon)} = \f{VL}{2G_N^{(5)}}\int_{r_0}^{r_\infty}dr \f{r}{R}
    = \f{VL}{4G_N^{(5)}R}(r_\infty^2 - r_0^2)\ .
\end{align}

\begin{figure}
\begin{center}
\includegraphics[width=6cm,clip]{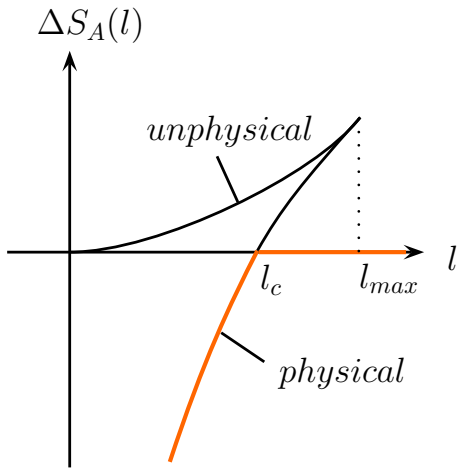}
\end{center}
\caption{
\label{fig:entropydiff}
The entanglement entropy as a function of the width $l$. There are three solutions which are locally minimal
area surfaces: two connected surfaces and a disconnected one. We set $\Delta S_A=0$ for the disconnected one.
One of the connected one has larger area than the
other and is unphysical. When $0<l<l_{c}$ the connected one is chosen, while when $l>l_c$ the disconnected one
becomes dominant.}
\end{figure}

We plot the difference of the entanglement entropy between the connected and disconnected surfaces $\D S_A \equiv S_A^{(con)} -
S_A^{(discon)}$ as a function of the length $l$ of the subsystem $A$ in
Fig.\ \ref{fig:entropydiff}.
Notice that the physical solution in Fig.\ \ref{fig:entropydiff} (i.e. lower branch) is
concave as a function of $l$, being consistent with the strong subadditivity of the von-Neumann entropy
(see Sec.\ \ref{prop EE}).
When $\D S_A$ becomes positive at the critical length $l_c(<l_{max})$, the disconnected surface dominates, i.e.,
becomes minimal. Then there happens a phase transition at $l=l_c$, which corresponds to the confinement/deconfinement
phase transition in dual gauge theory \cite{Nishioka:2006gr,Klebanov:2007ws}.

A similar analysis has been done in \cite{Klebanov:2007ws} in the more general backgrounds
including the Klebanov-Strassler solution \cite{KlSt}.
These results indicate that
the entanglement entropy can be a good order parameter for a phase transition.
A benefit of the holographic entanglement entropy is that in order to detect
the confinement/deconfinement, we do not need finite temperature black brane solutions,
which are often difficult to get analytically.

In summary, our holographic analysis predicts the following behavior of the finite part of the
entanglement entropy in $(d+1)$-dimensional
confining large $N$ gauge theories
at vanishing temperature (we subtracted the area law divergence $\sim a^{-(d-1)}$):
\ba
&& S_A(l)|_{finite}=-VF(l)\ ,\no
&&\mbox{where}\ \ \ \  F(l)\simeq c_1 N^2 l^{-(d-1)} \ \ \ (l\to 0)\ ,\no
&& \ \ \ \ \ \ \ \ \ \ \ \,   F(l)= c_2 N^2 \ \ \ (l>l_c)\ , \label{phasetq}
\ea
where $V$ is the volume of the non-compact $d-2$ directions transverse to the separation.
The numerical coefficients $c_1$ and $c_2$ depend on each theory.

Remarkably, the numerical computation of the entanglement entropy in the lattice gauge theory
has been done in \cite{Velytsky:2008rs,Buividovich:2008kq,Buividovich:2008gq,Buividovich:2008yv}, and
the non-analytic behavior like Fig.\ \ref{fig:entropydiff} or (\ref{phasetq}) has been confirmed.
These results would also support the validity of the holographic formula (\ref{arealaw}) of
the entanglement entropy in the AdS/CFT correspondence.

\begin{figure}
\begin{center}
\includegraphics[width=5cm,clip]{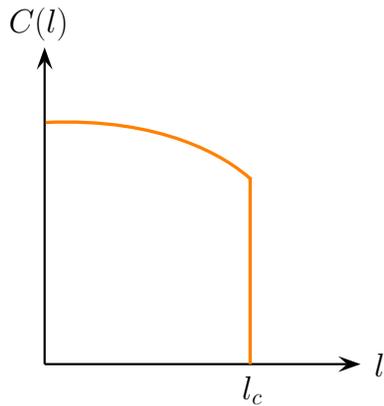}
\end{center}

\caption{\label{fig:entropic_c}
The entropic c-function with respect to $l$.
It jumps to zero at $l=l_c$, which is identified with
the confinement/deconfinement transition.}
\end{figure}

To make the phase transition clear,
we can define the following quantity called an ``entropic c-function''
\begin{align}
    C(l) \equiv \f{l^{d}}{V}\f{dS_A(l)}{dl} \ ,
\end{align}
which does not depend on the UV cutoff. This is a natural generalization
of the entropic c-function defined in two dimensions \cite{Casinicth,Casini:2006es}.
 We observe that $C(l)$ is a monotonically decreasing
function of $l$, which is regarded as a entropy version of the c-theorem as
$C(l)$ measures the degrees of freedom at the energy scale $\Lambda \sim 1/l$.
Its explicit form is sketched in Fig.\ \ref{fig:entropic_c}.
The sharp dump of $C(l)$ is because the AdS bubble solution is completely cutoff
in the IR region $r< r_0$ and represents the mass gap in dual gauge theory. For finite $N$ gauge theories,
the behavior may become milder.

Finally, it is also intriguing to mention
a relation
to closed string tachyon condensation. It is argued that in \cite{HoSi} that the
AdS soliton corresponds to the state after closed string tachyon condensation in the background of
compactified $\mathrm{AdS}_5$ with anti periodic boundary conditions for fermions.
In this interpretation, we observe that the entanglement
entropy is decreased after the tachyon condensation like the ADM energy \cite{Energy} of the
background \cite{Nishioka:2006gr}. This might suggest that the
entanglement entropy is an important quantity which characterizes the closed string tachyon condensation.

\subsection{Geometric Entropy}\hspace{5mm}
In the previous sections, we always assumed the vanishing temperature
and we found that the entanglement entropy can be an order parameter
for a phase transition in confining gauge theory.  It is also interesting to
detect the confinement/deconfinement phase transition at finite temperature. Unfortunately,
the entanglement entropy cannot probe the thermal phase transition because it is
defined at the specific time and does not wind the thermal cycle
(refer to \cite{Faraggi:2007fu} for detailed calculations).
Instead, we can define the geometric entropy, which is regarded as the double Wick rotated
version of the entanglement entropy \cite{Fujita:2008zv}.
The relation between the geometric entropy and the ordinary entanglement entropy is
analogous to the one between the Polyakov loop and the Wilson loop.

To illustrate the definition of the geometric entropy, we consider the gauge theory
on $S^3$ at finite temperature. We express the metric of $S^3$ as follows
\begin{align}
    d\Omega_{(3)}^2 = d\t^2 + \sin^2\t (d\psi^2 + \sin^2\psi d\phi^2) \ ,
\end{align}
where $0 \le \t,\, \psi \le \pi$ and $0 \le \phi \le 2\pi$.
If we change the periodicity of $\phi$ to $0\le \phi \le 2\pi/n$,
there exists
conical singularities at $\psi= 0$ and $\psi =\pi$ with the deficit angle $\d = 2\pi (1-1/n)$.
Then the gauge theory is defined on the orbifold $S^3/Z_n$.
Considering the partition function on the orbifolded space, we can define the geometric
entropy following the usual definition of the
von-Neumann entropy \cite{Fujita:2008zv}
\begin{align}
    S_G = -\f{\p}{\p (1/n)}\log \left[ \f{Z_{YM}(S^3/Z_n)}{(Z_{YM}(S^3))^{1/n}} \right] \Bigg|_{n=1} \ .
\end{align}

If we have a dual geometry for the gauge theory, we can also perform the holographic calculation
of the geometric entropy. When we require the boundary should be $S^1 \times S^3$, we have two solutions
in the bulk space \cite{Witten}: one is the thermal AdS space
and the other is the Schwarzschild AdS black hole. Using the bulk to boundary relation in the supergravity approximation, we obtain
the holographic formula for the geometric entropy similarly to (\ref{HolEE})
\begin{equation}
S_G =\frac{\mbox{Area}(\gamma)}{4G_N^{5}} \ ,
\end{equation}
where the surface $\g$ is defined by $\sin\psi =0$ in the bulk and it winds the thermal cycle $\tau$.

It is well-known that there is a thermal phase transition between the thermal AdS space
and the Schwarzschild AdS black hole, called the Hawking-Page transition \cite{HaPe},
when we change the temperature or the period of the thermal cycle $\b$.
From the viewpoint of the AdS/CFT correspondence, there should be a corresponding phase transition
in dual CFT on $S^3$ \cite{Witten}, and actually there exists a confinement/deconfinement phase
transition in it \cite{Su,AMMPR}.
The geometric entropy can capture this phase transition in both gravity and gauge theory sides
and then, this quantity can be a useful order parameter for a confinement/deconfinement transition at
finite temperature. Indeed, the result in the gravity side qualitatively agrees with that in the free Yang-Mills theories
as shown in \cite{Fujita:2008zv}.
See \cite{Bah:2008cj} for the application of the geometric entropy as an order parameter to the other backgrounds.

\subsection{Topological Entanglement Entropy and Boundary Entropy}
\hspace{5mm}
As we mentioned, in a gapped system whose low energy theory is described by a topological field theory,
the entanglement entropy offers us important information about the ground state.
This kind of systems are rather
common in $(2+1)$-dimensional condensed matter systems. For example, the quantum Hall effect occurs in materials whose
low energy theory is described by an abelian Chern-Simons gauge theory (see e.g. \cite{Wen89}).
The appearances of non-abelian
Chern-Simons theories have also been discussed in a similar context \cite{MooreRead}.

In such $(2+1)$-dimensional systems with a mass gap, the entanglement entropy takes the form (assuming $A$ is
a disk)
\be
S_A=\gamma\cdot\f{l}{a}+S_{top}\ ,
\ee
where the first term in the right-hand side is the area law divergence. The second term $S_{top}$ is a finite
quantity and is called the topological entanglement entropy.
In \cite{Levin05,Kitaev05}, this has been shown to be invariant under any smooth deformations of the subsystem $A$ and $S_{top}$
has been calculated explicitly.
The calculations based on the surgery method in Chern-Simons gauge theory \cite{WiCS}
have been performed in \cite{DFLN}.

Therefore, it is intriguing to calculate $S_{top}$ holographically. In the absence of
the Chern-Simons term,
it has been calculated in \cite{Pakman:2008ui} for the pure Yang-Mills in $2+1$
dimensions
and found that
$S_{top}=0$, which is consistent with the gauge theory side. To obtain non-trivial results we need to
include the Chern-Simons interaction. In \cite{Fujita:2009kw}, it has been clarified how the expected
result of $S_{top}$ can be holographically obtained by considering a D3-D7 system and by treating D7-branes
as probes. A direct supergravity computation of $S_{top}$ is still a future problem.

It is also intriguing to note that in these topologically ordered systems,
there is a precise connection between physics in the bulk of the system
and at the boundaries
(this correspondence is sometimes called `holography' in condensed matter physics.)
\cite{Fendley:2006gr}.
For example,
in the Chern-Simons gauge theory, information in the bulk,
such as the fractional charge and statistics of quasi particle excitations,
can be mapped to
chiral conformal field theory which is realized at
the (1+1) dimensional edge (boundary) of the (2+1) dimensional system.
In this correspondence, the topological entanglement entropy is
equal to the boundary entropy in the conformal field theory \cite{Fendley:2006gr}.
In the language of AdS/CFT, we can indeed realize this
bulk/edge duality as the AdS$_3$/CFT$_2$ correspondence \cite{Fujita:2009kw}.

The boundary entropy is originally defined as the ground state degeneracy due to the presence of
boundary in two-dimensional CFTs \cite{AfLu}. Actually, it also coincides with the finite part of
the entanglement entropy which arises due to the presence of boundary \cite{Cardy} (for a review see 
\cite{Afl}). By using this relation,
a holographic calculation of boundary entropy has successfully
been done in \cite{Azeyanagi:2007qj} based on AdS$_3$/CFT$_2$.

\section{BH Entropy as Entanglement Entropy}
\setcounter{equation}{0}
\label{bfrs}
\hspace{5mm}
An important original motivation for the entanglement entropy
in quantum field theories has been the
microscopic understanding of the black hole entropy. Even though
the entropy of supersymmetric (BPS) black holes has been understood by
explicitly counting the BPS states \cite{StVa}, the entropy of Schwarzschild black holes
has not been well understood microscopically. It is natural that
some sort of quantum entanglement
between
the inside and outside of the event horizon is relevant for the explanation of the entropy of
the Schwarzschild black hole. Indeed, as we have explained in Sec.\ \ref{EEBH},
the entanglement entropy shares
similar properties with the black hole entropy.

In the case of
induced gravity, the entanglement is essentially equivalent to
the black hole entropy
(see e.g.\cite{Jacobson:1994iw}). Interestingly, we can confirm this holographically by
 considering an analogue of AdS/CFT in the brane-world setup (RS II \cite{RS}) \cite{HMS,Emparan:2006ni,Iwashita:2006zj,Solodukhin:2006xv} as we review in Sec.\ \ref{secbh}.

There is another way to relate the black hole entropy to the entanglement entropy.
This is given by directly applying AdS/CFT
to AdS black holes. The most well-known example
is the AdS-Schwarzschild solution. It is clearly dual to a CFT at finite temperature. At the same time,
we can start with a pure state (Hartle-Hawking state) in a pair of
these CFTs (called CFT1 and CFT2)
\be
|\Psi\lb=\f{1}{\s{Z}}\sum_{n}e^{-\beta E_n/2}|n\lb_1\otimes |n\lb_2\ ,
\ee
where $E_n$ denote energy eigenvalues of the given CFT and we defined $Z=\sum_{n}e^{-\beta E_n}$.
Indeed, by
tracing out one of
the Hilbert space of the 2nd CFT
we get correctly the thermal density matrix:
\be
\rho_1=\mbox{Tr}_2|\Psi\lb \la \Psi|=\f{1}{Z}\sum_n
e^{-\beta E_{n}}
|n\lb_1 \la n|_1\
.
\ee
This happens exactly
in the AdS-Schwarzschild black hole since its extended Penrose diagram
has two boundaries, which are identified with
CFT1 and CFT2
as found in \cite{MBH}.
In Sec.\ \ref{secbhh} below, we will explain that a similar interpretation is also possible for
a pure AdS$_2$ space and this enables us to understand the entropy of extremal black holes in
flat spacetimes as the entanglement entropy of certain systems of conformal quantum mechanics
assuming AdS$_2$/CFT$_1$ \cite{Azeyanagi:2007bj} (see also relevant discussions in \cite{Carroll:2009ma}).
For other recent progresses
on the relation between
the black hole entropy and entanglement entropy, refer to
\cite{BEY,Brustein:2006wp,Cadoni:2007nh,Cadoni:2007vf,Casini:2007dk,Das:2007mj,Das:2008sy}.

\subsection{BH Entropy as Entanglement Entropy via Brane-World}\label{secbh}
\hspace{5mm}
Let us remember that the AdS/CFT correspondence with a UV cut off $z>a$ can
be regarded as a brane-world setup (RS2 \cite{RS}).
In this context, we usually generalize AdS/CFT
so that the cut off $a$ to be of order $R$ (AdS radius) and
a Newton constant $G^{brane}_{N}\sim\f{d-1}{R}G^{bulk}_{N}$  is induced on the brane. By assuming the extension
of AdS/CFT to this system, we find that the $(d+1)$-dimensional quantum gravity on the brane
is dual to the classical gravity on the $(d+2)$-dimensional AdS space with the cut off.
This description offers us an interesting way to treat a black hole including
quantum corrections \cite{EHM}.

If one wants to work within the standard AdS/CFT conservatively, we can assume that the cut off $a$ is small
$a\ll R$. Then the Newton constant on the brane becomes very small
\begin{equation}
  \frac{1}{G^{brane}_N}\sim \frac{R^d}{G^{bulk}_{N}}\int^\infty_a
    \frac{dz}{z^{d}}=\frac{R^d}{(d-1)a^{d-1}}\frac{1}{G^{bulk}_{N}}\gg
    \frac{R}{G^{bulk}_{N}}\ .  \label{weakg}
\end{equation}
This weak gravity system is also enough for our purpose below.

In the paper \cite{EHM}, authors construct four-dimensional black
hole solutions to the vacuum Einstein equation with the negative
cosmological constant (see also \cite{Anber} for further analysis).
The horizon $\Sigma$ extends toward the $(2+1)$-dimensional brane
and the induced
 metric on the brane looks like
Schwarzschild metric\footnote{On the brane, we expect no cosmological constant.
In usual Einstein gravity with zero cosmological constant, there is no black hole solution.
In our case, the result should be interpreted such that it already contains quantum corrections following
the philosophy of AdS/CFT. We expect that large quantum corrections make such a black hole solution
possible \cite{EHM}.}
\be
ds^2_{brane}=-\left(1-\f{r_0}{r}\right)dt^2+\f{dr^2}{1-r_0/r}+r^2
d\phi^2\ .
\ee
In the middle of the bulk $\mathrm{AdS}_4$,
the size of the horizon shrinks to zero
and thus its topology is a disk, which looks very similar to
the setup (b) in Fig.\ \ref{fig: min_surf}. Recently, brane world black hole for
$\mathrm{AdS}_5$ has been obtained
in \cite{KaRe} by considering extremal brane-world black holes with the
$\mathrm{AdS}_2\times S^2$ near
horizon geometry.

Now we would like
to apply the holographic entanglement entropy to brane-world black holes.
Let us choose the subsystem $A$ is inside the horizon $r=r_0$ on the brane. Then the minimal surface $\gamma$
which is the bulk extension of $A$ is actually given by the horizon $\Sigma$ of the bulk black hole
solution \cite{EHM}. Thus we find that the holographic entanglement entropy (\ref{arealaw}) coincides with the
Bekenstein-Hawking entropy of the $\mathrm{AdS}_4$ black hole solution. The latter is considered to be
equal to the quantum corrected black hole entropy of the $(2+1)$-dimensional brane-world black hole via AdS/CFT.
Therefore, we can conclude that the entropy for the entanglement between the inside and outside of the horizon is
the same as the black hole entropy with quantum corrections \cite{Emparan:2006ni,HMS}.

Notice that the classical entropy $\f{Area(\Sigma)}{4G^{brane}_{N}}$ in the brane gravity largely deviates from the quantum corrected one
$\f{Area(\gamma)}{4G^{bulk}_{N}}$ when $a\sim R$.
We can see the above claim explicitly by computing the holographic entanglement entropy.
Assuming that $a$ is very small, the entanglement entropy $S_A$ behaves like
\be
S_A=\f{\mbox{Area}(\gamma)}{4G^{bulk}_N}=\gamma\f{\mbox{Area}(\Sigma)}{a^{d-1}}+O(a^{d-2})=
\f{\mbox{Area}(\Sigma)}{4G^{brane}_{N}}+O(a^{d-2})\ ,\label{entbhre}
\ee
where $\gamma$ is a certain numerical factor. The subleading term $O(a^{d-2})$ can be interpreted as
the quantum corrections to the classical Bekenstein-Hawking formula. In this weak gravity limit $a\to 0$, the leading term becomes dominant. It is amusing to note that in (\ref{entbhre}), the area law term of the entanglement
entropy in quantum field theories essentially becomes equal to the black hole entropy.
These arguments strongly suggest that some sort of induced gravity is realized in the brane-world setup.

A similar interpretation of two-dimensional black holes has been found in \cite{Solodukhin:2006xv}.
We can also apply the same brane-world argument to explain the entropy of de-Sitter spacetime as
discussed in \cite{HMS,Iwashita:2006zj}.

\subsection{BH Entropy as Entanglement Entropy via
  AdS$_2/$CFT$_1$}\label{secbhh}
\hspace{5mm}
The pure AdS spacetime AdS$_{d+1}$ with $d\ge 2$ has no entropy as is
also clear from its dual CFT$_d$ at zero temperature.
To obtain non-zero entropy, we need to consider the AdS black hole as
the dual geometry. On the other hand, we expect non-zero entropy for the
pure AdS$_2$ spacetime since it appears as the near horizon limit of higher-dimensional extremal black holes
\cite{Astefanesei:2006dd,Kunduri:2007vf,AY,Kunduri:2008rs,Kunduri:2008tk}.
Thus the microscopic interpretation of the Bekenstein-Hawking entropy of
the extremal black holes would be related to the AdS$_2$/CFT$_1$
correspondence \cite{Strominger:1998yg,HaSt}. Even though
the AdS$_2/$CFT$_1$ has not been well understood as opposed to the higher-dimensional AdS/CFT, below we assume that the gravity on $AdS_2$ is dual to a certain
conformal quantum mechanics (CFT$_1$). In other words, one may think that the following argument
is an indirect evidence for AdS$_2/$CFT$_1$.
A formulation based on
the entropy function has been done in \cite{Sen}. Also the
the appearance of
the AdS$_2$ spacetime plays an important role in the recent investigations
of the attractor mechanism (see \cite{Sen:2007qy} and references therein), and
more recently, in a new duality called the extremal black hole/CFT
correspondence \cite{Guica:2008mu,Hartman:2008pb}.

The AdS$_2$ geometry has a special property such that it has two
timelike boundaries in the global coordinate
\begin{align}\label{AdSmet}
 ds^2 = \ell^2 \frac{-d\tau^2 + d\sigma^2}{\cos^2\sigma} \ ,
\end{align}
where $\ell$ is the radius of the AdS space and $-\f{\pi}{2} \le \sigma \le \f{\pi}{2}$.
Then, according to the principle of AdS/CFT, we expect that there are
two CFT$_1$s on the
boundary $\sigma = \pm \f{\pi}{2}$ of the AdS$_2$ space.
Here we would like to show that the black hole entropy is exactly the
same as the entanglement entropy between the two CFTs by using the AdS$_2$/CFT$_1$
correspondence. Actually we can show that the two CFTs are entangled
applying the holographic formula of the entanglement entropy (\ref{RTF}) as
\begin{align}\label{entadst}
 S_{ent}=\f{\mbox{Area}(\g_A)}{4G^{(2)}_N}=\f{1}{G^{(2)}_N} \ .
\end{align}
This is because the minimal surface now becomes
a point. Below we will give a clearer derivation of (\ref{entadst})
based on AdS/CFT \cite{Azeyanagi:2007bj}.

As we mentioned above, there are two independent CFTs on the boundaries
of the AdS$_2$ space, namely CFT1 and CFT2.
The Hilbert spaces of CFT$1$ and CFT$2$ are denoted by $H_1$ and
$H_2$. The total Hilbert space looks like $H_{tot}=H_1\otimes H_2$.
We define the reduced density matrix from the total density matrix
$\rho_{tot}$
\begin{align}
 \rho_1=\mbox{Tr}_{H_2}\rho_{tot} \ ,
\end{align}
by tracing over the Hilbert space $H_2$.
This is the density matrix for an observer who is blind to CFT$2$.
It is natural to assume that $\rho_{tot}$ is the one for a pure state.

\begin{figure}[t]
\begin{center}
  \hspace*{0.5cm}
  \includegraphics[height=5cm]{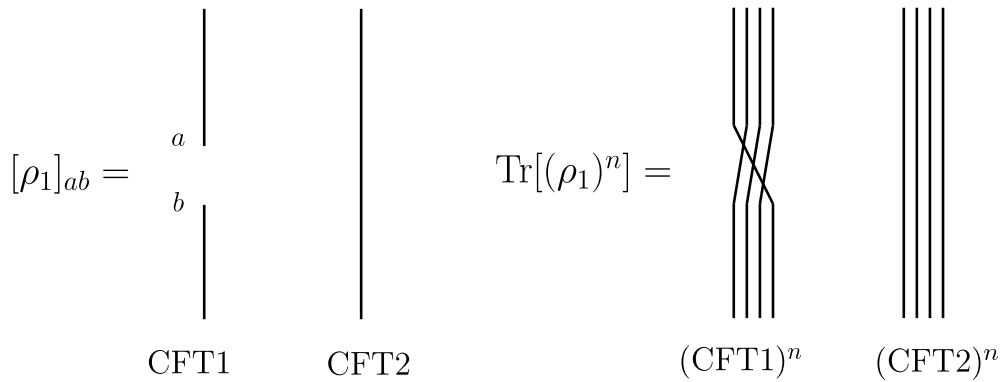}
  \caption{The calculation of reduced density matrix $\rho_1$}\label{qft}
\end{center}
\end{figure}

The entanglement entropy for CFT$1$, when we assume that the
opposite part CFT$2$ is invisible for the observer in CFT$1$, is
defined by
\begin{align}
 S_{ent}=\mbox{Tr} [-\rho_1 \log \rho_1] \ .
\end{align}
We can obtain this by first computing Tr$(\rho_1)^n$, taking the
derivative w.r.t.$~n$ and finally setting $n=1$. In the path
integral formalism of the quantum mechanics, $\rho_1$ and
Tr$(\rho_1)^n$ are computed as in Fig.\ \ref{qft} (we perform the
path-integral along the thick lines and $a$ and $b$ are the
boundary conditions).

\begin{figure}[t]
\begin{center}
\hspace*{1cm}
  \includegraphics[height=6cm]{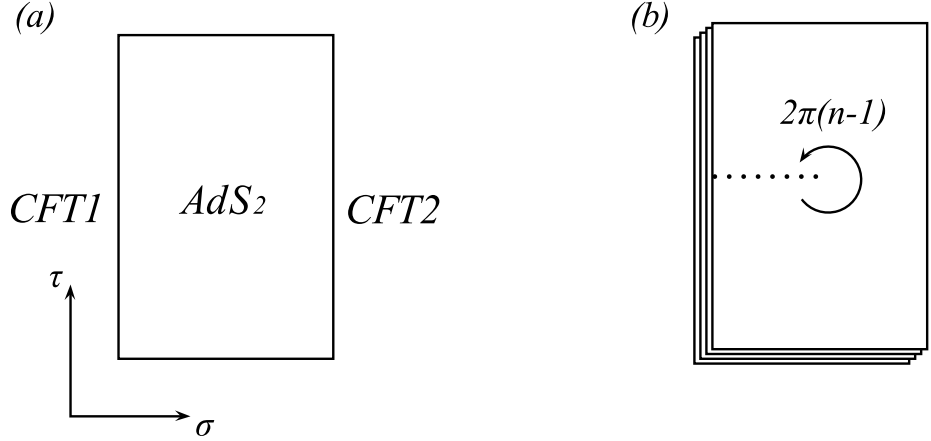}
  \caption{The geometry of AdS$_2$ with two boundaries [Left] and the 2D
 spacetime ($n$-sheeted Riemann surface with a cut) which is dual to the
 computation Tr$(\rho_1)^n$ [Right].}\label{adstwo}
\end{center}
\end{figure}

By using the bulk-boundary relation of AdS/CFT \cite{ADSGKP,ADSWitten}, we can
compute the entanglement entropy holographically\footnote{
Our derivation seems to be closely related to the conical defect argument of black hole entropy
(see e.g. \cite{Jacobson:1994iw,FursaevR}). However, notice that in these arguments the authors consider
the entanglement entropy for the total spacetime of
non-extremal black holes, while in our argument we consider the
entanglement entropy for the boundary of the extremal black
hole geometry. See also \cite{Carroll:2009ma} for a discussion on the relevance of the
$\mathrm{AdS}_2$
geometry.} as in the right panel of
Fig.\ \ref{adstwo}. The dual geometry is the $n$-sheeted
Riemann surface \cite{RuTa,RuTaL}, assuming the Euclidean metric.
The cut should end at
a certain point in the bulk
because there should not be any cut on the opposite boundary, which
is first traced out. Notice that the presence of two boundaries in
AdS$_2$ plays a crucial role in this holographic computation. We
would get the vanishing entropy if we were to start with the
spacetime which has a single boundary such as the Poincare metric of
AdS$_2$.

Now we remember the Einstein-Hilbert action in the Euclidean space
\begin{align}
 I =-\f{1}{16\pi G^{(2)}_N}\int dx^2 \s{g}(R+\Lambda) \ .
\end{align}
The cosmological constant $\Lambda$ is not important since
the contribution to the Einstein-Hilbert action
from the cosmological constant term is
extensive and it will vanish in the end of the entropy computation.
In the $n$-sheeted geometry we find $I=\f{n-1}{4G^{(2)}_N}$ in
the Euclidean formalism because the curvature behaves like a delta
function $R=4\pi (1-n)\delta^2(x)$ (see e.g.\cite{Fursaev:2006ih,FursaevR}). The
entanglement entropy is obtained as follows
\begin{align}
 S_{ent}=-\f{\de}{\de n}\log ( e^{-I + nI^{(0)}}
)|_{n=1} =\f{1}{4G^{(2)}_N} \ ,
\end{align}
where $I^{(0)}$ is the value of the Einstein-Hilbert action of a
single-sheet in the absence of the cut (or negative deficit angle).

Recently, it has been shown that extremal (rotating) black holes
always have the $SO(2,1)$ symmetry in the near horizon limit
\cite{Astefanesei:2006dd,Kunduri:2007vf,AY,Kunduri:2008rs,Kunduri:2008tk}.
For example, the near horizon geometry of a four-dimensional extremal Kerr black hole is given by a warped product of
AdS$_2$ and a two-dimensional manifold \cite{BaHo}.
Then we can assume the near horizon geometry of $d$-dimensional extremal black holes as
\begin{align}
 ds^2 = fds^2_{AdS_2} + ds^2_{M^{d-2}} \ ,
\end{align}
where the first term is the AdS$_2$ space in the Poincare coordinate
$ds^2_{AdS_2} = -\frac{r^2}{\ell^2}dt^2 + \frac{\ell^2}{r^2}dr^2$ with a warp factor $f$
that depends on the coordinate of $M^{d-2}$; the
second term is the metric of the compact manifold $M^{d-2}$ of the horizon such as $S^{d-2}$.
The horizon is at $r=0$ in this coordinate and the Bekenstein-Hawking
entropy is
\begin{align}
 S_{BH}= \frac{\text{Vol}(M^{d-2})}{4G_N^{(d)}} \ .
\end{align}
Finally, it is trivial to see that
\begin{align}
 S_{ent}=S_{BH} \ ,
\label{entbb}
\end{align}
because the Newton constant in two dimensions is defined as $\frac{1}{G^{(2)}_N}\equiv
\frac{\text{Vol}(M^{d-2})}{G^{(d)}_{N}}$.
This means that the entanglement between
CFT$1$ and CFT$2$ is precisely the source of the extremal
black hole entropy.

Moreover, we can take curvature corrections into
account. We assume that the near horizon geometry is of the form
AdS$_2 \times M^{d-2}$ even in the presence of the higher derivative corrections.
Even though we start with the Lagrangian
${\cal L}$ that includes the curvature tensor $R_{\mu\nu\rho\sigma}$
and their covariant derivatives, we can neglect the covariant
derivative of curvature tensors because the near horizon geometry
has the constant curvature. In this case, the black hole entropy
with the curvature corrections is given by the Wald's formula
\cite{Wa,IyWa,JKM}
\begin{align}
 S_{BH}=-2\pi \int_{{\cal H}}\s{h}\f{\de
{\cal{L}}}{\de R_{\mu\nu\rho\sigma}}\ep_{\mu\nu}\ep_{\rho\sigma}\ ,
\end{align}
where $\ep_{\mu\nu}=\xi_\mu\eta_\nu-\xi_\nu\eta_\mu$
by using
the Killing vector $\xi_\mu$ of the Killing horizon and its normal
$\eta_\nu$, normalized such that $\xi\cdot \eta=1$; ${\cal H}$
represents the horizon and $h$ is the metric on it.
Reducing the $d$-dimensional metric to AdS$_2$ space, the action might change
into that with higher derivative corrections and non-gravitational fields
such as gauge fields. Fortunately, the non-gravitational
fields do not contribute to the Wald's formula and we can neglect these
terms and concentrate on the higher derivative action even in two dimensions.

Now we would like to compare the Wald entropy with the entanglement
entropy computed holographically via AdS$_2$/CFT$_1$. We consider
the $n$-sheeted AdS$_2$, where the
Riemann tensor behaves as follows \cite{FursaevR}
\begin{align}
R_{abcd}=R_{abcd}^{(0)}+2\pi(1-n)\cdot
(g_{ac}g_{bd}-g_{ad}g_{bc})\cdot \delta_H\ .
\end{align}
Here $\delta_H$ is the delta function localized at the (codimension two)
horizon (the $H$ is actually a point in AdS$_2$ and is related to the
original horizon $\cal H$ as ${\cal H} = H \times M^{d-2}$).
$R_{abcd}^{(0)}$ represents the constant curvature contribution from
the cosmological constant. $a,b$ run the coordinate in the AdS$_2$.
Notice also that if we employ the relation
$g_{ab}=\xi_a\eta_b+\xi_b\eta_a$, we obtain
$\ep_{ab}\ep_{cd}=-(g_{ac}g_{bd}-g_{ad}g_{bc})$.
Now we consider the perturbative expansions of the Lagrangian with
respect to the (delta functional) deviation of $R_{abcd}$ from
$R_{abcd}^{(0)}$. Then the quadratic and higher order terms do not
contribute since $\lim_{n\to 1}\f{d}{dn}(1-n)^p=0$ for $p\geq 2$.
Therefore, we can find
\begin{align}
 I_n=-\log Z_n
 =& \, 2\pi (1-n)\int_{H} \s{h}\f{\de {\cal{L}}}{\de
R_{abcd}}\ep_{ab}\ep_{cd} \ .
\end{align}
Thus this agrees with the Wald's formula in two dimensions
\begin{align}
S_{ent}=-\f{\de}{\de n}\log Z_n\Bigl |_{n=1}=-2\pi\int_{H}\s{h}\f{\de
{\cal{L}}}{\de R_{\mu\nu\rho\sigma}}\ep_{\mu\nu}\ep_{\rho\sigma}=S_{BH} \ .
\end{align}
After introducing the 2D Newton constant as before, the Wald entropy in
two dimensions is exactly the same as that in $d$ dimensions, and we
complete the proof of the equivalence of the black hole entropy and
the entanglement entropy even in the presence of the higher derivative
correction.

\section{Covariant Holographic Entanglement Entropy}\label{seccov}
\setcounter{equation}{0}
\hspace{5mm}
So far we have only discussed static spacetimes. It is straightforward to
extend our holographic formula to static spacetimes which are not asymptotically
AdS as long as we have its holographic dual theory.
However, it may be
more interesting to consider holography in a time-dependent
spacetime as eventually we would like to understand cosmological
backgrounds such as the de-Sitter space from a holographic
viewpoint.

\subsection{Covariant Entropy Bound}
\hspace{5mm}
In the previous argument of Sec.\ \ref{holographic}, we assumed a time slice on which we can
define minimal surfaces since its signature is Euclidean. However,
in the time-dependent case there is no longer a natural choice of
the time-slices as we have infinitely many different ways of
defining the time slices. Thus we need to consider the entire
Lorentzian spacetime. Then we are in trouble
since in Lorentzian
geometry there is no minimal area surface as the area vanishing if
the surface extends in the light-like direction. In order to resolve
this issue, let us remember an analogous problem; the covariant
entropy bound so called the Bousso bound \cite{Bousso}.

In general, if we get heavy objects together in a small region and
continue to bring another one into the region, this system
eventually experiences the gravitational collapse. Therefore we have
an upper bound of the mass and entropy which can be included inside
of the surface $\Sigma$. The bound for the entropy in flat space
time is called the Bekenstein bound and it is given by
\begin{equation}\label{Bek}
S_{\Sigma}\leq \frac{\mbox{Area}(\Sigma)}{4G_N}\ ,
\end{equation}
 where $\Sigma$ is a
codimension two closed surface in the spacetime. It is also more
interesting to generalize this bound to any time-dependent
backgrounds like the cosmological ones. This requires to find a
covariant description. It is obvious that the Bekenstein bound
(\ref{Bek}) is not covariant since the definition of the entropy
included inside $\Sigma$ is not covariant but depends on the choice
of the time slice. The covariant entropy bound was eventually
formulated by Bousso \cite{Bousso} and it is given by
\begin{equation}\label{bousso}
S_{L(\Sigma)}\leq \frac{\mbox{Area}(\Sigma)}{4G_N}\ .
\end{equation}
The light-like manifold $L(\Sigma)$ is called the light-sheet of
$\Sigma$. This is defined by the manifold which is generated by the
null geodesics starting from the surface $\Sigma$. We require that
the expansion $\theta$ of the null geodesic is non-positive
$\theta\leq 0$. In the flat spacetime, this is just a half of
light-cone and the same is true for the AdS spacetime as it is
conformally flat. Then the quantity $S_{L(\Sigma)}$ means the
entropy which passes
through the light sheet $L(\Sigma)$, which is
covariantly well-defined.
One more interesting thing on
the Bousso bound is that we can apply the bound even if the surface $\Sigma$
has boundaries, which is quite useful in the holographic setup as we
employ below.

\subsection{Covariant Holographic Entanglement Entropy}
\hspace{5mm}
Now we would like to return to our original question of the
covariant holographic entanglement entropy. Our final claim
\cite{Hubeny:2007xt} is given by
\begin{equation}\label{cov}
    S_{A}(t)=\frac{\mbox{Area}(\gamma_{A}(t))}{4G^{d+2}_N}\ ,
\end{equation}
where $\gamma_A(t)$ is the extremal surface in the entire Lorentzian
spacetime ${\cal M}$ with the boundary condition $\partial
\gamma_A(t)=\partial A(t)$. The time $t$ is the time on the time
slice in the boundary $\partial {\cal M}=R^{1,d}$ and there is no unique way to
extend it to the bulk spacetime ${\cal M}$. This formula, for example, when applied to
 rotating BTZ black holes correctly
reproduces the entanglement entropy expected from CFT$_2$
\cite{Hubeny:2007xt}.

\begin{figure}
\begin{center}
\includegraphics[height=9cm,clip]{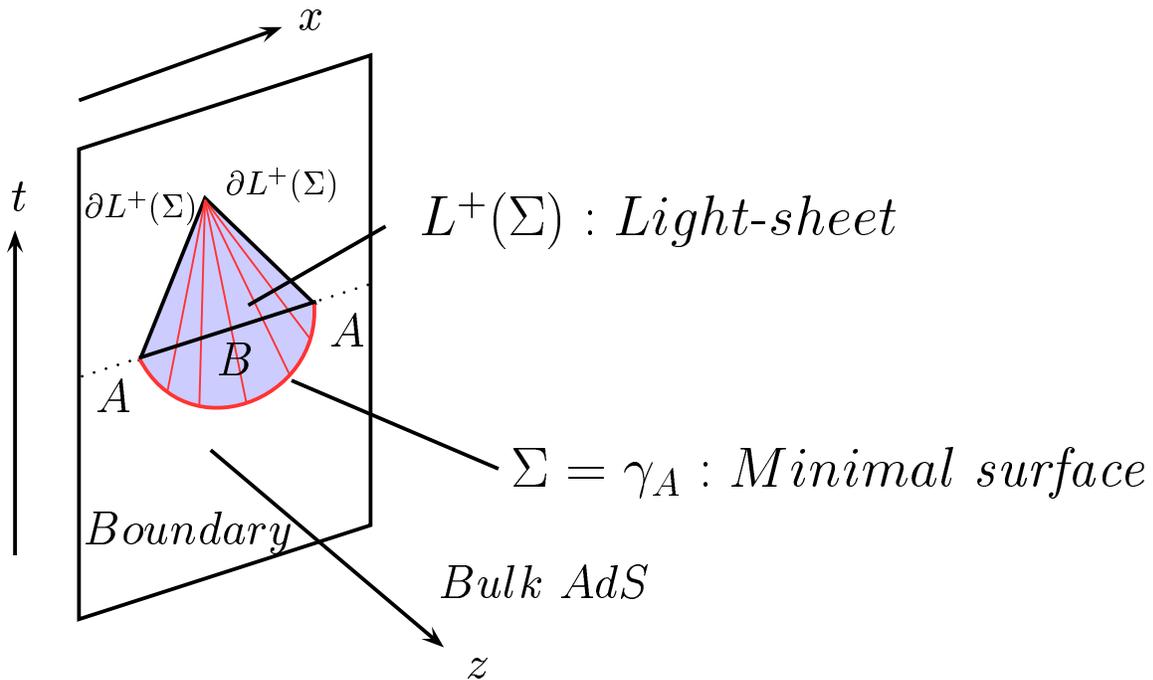}
\end{center}
\caption{ \label{bousso.eps} The setup of the Bousso bound applied to
the AdS$_3$/CFT$_2$ in the Poincare coordinate
$ds^2=\frac{R^2}{z^2}(-dt^2+dz^2+dx^2)$. In this simplified case,
the future Cauchy horizon coincides with the future
light-sheet $\partial L^+(\Sigma)$. In the above figure we only
write the future light-sheet and not the past one (i.e. $L^-(\Sigma)$)
just for simplicity.}
\end{figure}

This covariant formula (\ref{cov}) has originally been motivated
from the Bousso bound (\ref{bousso}) in \cite{Hubeny:2007xt}.
To see this let us again
remember the fact that the AdS/CFT correspondence with a UV cut off
$z>a$ can be regarded as a brane-world setup (RS2 \cite{RS}).
Assuming that the cut off is close to the UV $a\ll R$, the gravity on
the $(d+1)$-dimensional brane theory is very weak as in (\ref{weakg}).
In this setup, we would like to ask what is the Bousso bound on the brane
gravity theory (see Fig.\ \ref{bousso.eps} in the simplest case of
AdS$_3$/CFT$_2$). We expect that the brane theory with quantum
corrections taken into account is dual to the bulk gravity theory
which is classical, based on the standard idea of the AdS/CFT
correspondence. Therefore we argue that the quantum corrected Bousso
bound on the brane can be found as the classical Bousso bound on the
brane.

First we start with the setup of Bousso bound at the boundary
$\partial {\cal M}$. We pick up a (closed) surface $\partial \Sigma$ which
separates a time slice into the subsystems $A$ and $B$ such that
$\partial A=\partial \Sigma$.
See
Fig.\ \ref{bousso.eps}.
Now we define their light-sheets.
We consider both ones directed to the future and past
and
call them $\partial L^+(\Sigma)$ and $\partial L^-(\Sigma)$,
respectively.
The
reason why we put the symbol $\partial$ is that we are interested in
their bulk extensions $L^\pm(\Sigma)$. Again there are infinitely
many different ways of extending the boundary light-sheets toward
the bulk. We define the surface $\Sigma$ by the intersection
$L^+(\Sigma)\cap L^-(\Sigma)$. For each choice of such a $\Sigma$, we get
the Bousso bound (\ref{bousso}).

Here the condition of non-positive expansions of the null geodesics
on the light-sheets i.e. $\theta^\pm \leq 0$ comes into play. If
it were not for this condition, we could choose arbitrary $\Sigma$ and
take them to be light-like. However, the condition is rather
strong enough that the area of allowed $\Sigma$ takes a non-trivial
minimum and therefore we can define an analogue of the minimal
surface in this Lorentzian spacetime. The minimum of the area
corresponds to the most strict Bousso bound for a given boundary
surface $\partial \Sigma$ or equally the choice of the subsystem
$A$.

This minimum of the area occurs when the expansions on the two
light-sheets are both vanishing $\theta^\pm=0$. This condition is
actually equal to the statement that the surface $\Sigma$ is an
extremal surface again called $\gamma_A$, which is defined by the
saddle point of the area functional in the Lorentzian spacetime
\cite{Hubeny:2007xt}.

The final assumption is that the quantum Bousso bound on the brane
will be saturated by the entanglement entropy. This is because
 the entanglement entropy represents a thermal entropy
plus quantum corrections and it is defined by assuming that the
subsystem $B$ is completely smeared, which will be expected to lead
to the maximal entropy allowed in the region. If we assume this,
then we immediately reach the holographic entanglement entropy
formula (\ref{cov}).

In this way, we can extract the entropy in the dual
time-dependent background
by looking at our holographic entanglement entropy.
One may think that the opposite may be true: the extraction of metric from the information of
the entanglement entropy in CFT. For recent discussions in this direction refer to
\cite{Hammersley:2007ab,Bilson:2008ab}.

\subsection{Applications to Black Holes}
\hspace{5mm}
In black hole backgrounds, taking the limit where $A$ is the total space at the boundary, $S_A$ becomes
equal to the thermal entropy. In generic time-dependent backgrounds,
the thermal entropy is also time-dependent
and is clearly defined uniquely as the von-Neumann entropy of the thermal density matrix.
On the other hand, in the dual gravity side, it corresponds to the black hole entropy
which is obtained from the area of black hole horizon. However,
we have two candidates of horizon: event horizon
and apparent horizon. These two horizons are different in time-dependent spacetimes.
Our formula (\ref{cov}) of the holographic entanglement entropy selects the latter i.e.
the apparent horizon \cite{Hubeny:2007xt}. This is because the extremal surface condition
requires the vanishing of null expansion and this precisely match the definition of
apparent horizon. Notice also that the definition of event horizon is global.
Also our formula offers us to uniquely determine the
time slice which is requires to define the apparent horizon. Indeed, the recent works
\cite{ChYa,Figueras:2009iu} show that the area of apparent horizons behaves in a sensible way
as the thermal entropy, while that of even horizon not.

Finally, let us discuss an example where we can apply the
above covariant formula. We consider the AdS Vaidya solution (see e.g. \cite{Hub})
\begin{equation}\label{va}
ds^2=-(r^2-m(v))dv^2+2dvdr+r^2d\phi^2\ .
\end{equation}
This is the solution to the Einstein equation with the negative
cosmological constant in the presence of null matter whose EM tensor
looks like $T_{vv}=\frac{1}{2r}\frac{dm(v)}{dv}$. The null energy
condition requires $T_{vv}\geq 0$ and thus we find that $m(v)$ is a
monotonically increasing function of the (light-cone) time $v$.

This background is asymptotically
$\mathrm{AdS}_3$ and if we assume that
$m(v)$ is a constant, then it is equivalent to the static BTZ black
hole \cite{BTZ} with the mass $m$. Thus our background (\ref{va})
describes an idealized collapse of a radiating star in the presence
of negative cosmological constant. The dual theory is expected to be
a CFT in a time-dependent background. The time-dependence comes from
the time-dependent temperature. We can now apply the covariant
entanglement entropy formula (\ref{cov}) and in the end we find
\begin{equation}\label{res}
S_A(v)=\frac{c}{3}\left[\log
\frac{l}{a}+\frac{m(v)l^2}{6}+\cdot\cdot\cdot\right]\ ,
\end{equation}
as the expansion of small $m(v)$. The null energy condition
guarantees that this is a monotonically increasing function of time.
This shows that the entanglement entropy in this background is a
monotonically increasing function of time as is so in the second law
of the thermal entropy. We believe this behavior of
the entanglement
entropy in black hole formation processes is rather general.
However, we would like to stress that we are not claiming that the
entanglement entropy is always increasing. For example if we start
with the system with maximally entangled, the entanglement entropy
will decrease after a small perturbation due to the de-coherence
phenomenon.

\section{Conclusions}\label{seccon}
\setcounter{equation}{0}
\hspace{5mm}
In this article, we reviewed the recent progresses of holographic understanding of
the entanglement entropy,
starting from the main holographic formula (\ref{arealaw}).
We mainly employed the setup of AdS/CFT, though we can
straightforwardly extend our results to more general spacetimes with their
holographic duals. Even though the formula (\ref{arealaw}) has not been rigorously
proven, there have been many evidences so far. Including the ones which we did not discuss in this
article, we can list important evidences as follows:
\begin{itemize}
  \item The area law (\ref{divarea}) known in QFT can be easily reproduced holographically.
       The warp factor in the AdS space leads to the UV divergence
       of the dual CFT as explained in Sec.\ \ref{general proposal}.
  \item We can holographically derive the strong subadditivity
  (\ref{Sub}) in a very simple way as we reviewed in Sec.\ \ref{stsub}.
  \item We find perfect agreements between the AdS and CFT sides
  in the AdS$_3/$CFT$_2$ setup as in Sec.\ \ref{AdS3/CFT2}.
  \item In higher dimensions, it is not easy to
  calculate the entanglement entropy in QFTs analytically. Still we
  can show the semi-qualitative agreements between the CFT and AdS
  calculations as reviewed in Sec.\ \ref{AdS d+2/CFT d+1}.
  Moreover, for the logarithmic terms of the
  entropy we can show the precise agreement as its coefficient
  under the condition that the extrinsic curvature of $\de A$ is vanishing
  \cite{RuTaL}.
  \item In the example of $\CN =4$ Yang-Mills theory compactified
  on a Scherk-Schwarz circle dual to the AdS soliton, we can holographically detect
  a confinement/deconfinement transition as reviewed in Sec.\ \ref{secpha}.
  This qualitatively agrees both with the free Yang-Mills result
    and with recent lattice simulations. Moreover, in an almost supersymmetric limit,
   the holographic
    result of the entanglement entropy quantitatively agrees with the free Yang-Mills
    result \cite{Nishioka:2006gr}.
  \item
  In the presence of a horizon, the minimal surface $\gamma_A$
  tends to wrap the (apparent) horizon. Then the wrapped part gives
  an extensive contribution to the holographic entanglement entropy.
 This shows that our holographic formula generalizes the Bekenstein-Hawking
 entropy formula, as explained in Sec.\ \ref{general proposal}
  and Sec.\ \ref{secbh}.
  \item
  We can regard black hole entropy as the entanglement entropy
  by applying the idea of the holographic entanglement entropy to
  either brane-world black holes or AdS$_2/$CFT$_1$ as reviewed in
  Sec.\ \ref{bfrs}.

 \item The holographic formula is nontrivially consistent with the
  covariant entropy bound (Bousso bound) as explained in Sec.\ \ref{seccov}.

\end{itemize}

There are many future directions.
One of the most important future problems is to
derive the holographic entanglement entropy from the first principle of AdS/CFT
(i.e. bulk to boundary relation).  On the other hand,
to confirm the AdS/CFT correspondence from the
viewpoint of the entanglement entropy, we need to develop methods of calculations of
the entanglement entropy in quantum field theories.

Since the holographic entanglement entropy is expected to be
a universal observable
in holography, this quantity may be useful when we try to extend the holography
to other spacetimes such as de-Sitter spaces. The precise relevance of the entanglement entropy
to the understanding of black hole entropy has also been an interesting future problem. The fact that our
covariant formula is closely related to the Bousso bound may be a clue to this problem.

It is also intriguing to apply our holographic entanglement entropy to various condensed matter
systems. A complete holographic calculation of topological entanglement entropy for various $(2+1)$
dimensional systems with non-trivial topological orders will certainly be waited to be done in near future.

\vskip6mm
\noindent
{\bf Acknowledgments}

\vskip2mm

This preprint is based on the invited review article for a special issue ``Entanglement
entropy in extended quantum systems'' in Journal of Physics A. We would like to thank
P. Calabrese, J. Cardy and B. Doyon for the kind invitation.
It is a great pleasure to thank T. Azeyanagi, M. Fujita, M. Headrick, T. Hirata,
V. Hubeny, A. Karch, W. Li, M. Rangamani and D. Thompson for fruitful
collaborations on the related topics.
The work of TN is supported by JSPS Grant-in-Aid for Scientific Research No.\,19$\cdot$3589.
TN and TT are also supported by World Premier International Research Center Initiative
(WPI Initiative), MEXT, Japan.
SR thanks Center for Condensed Matter Theory at University of
California, Berkeley for its support.
The work of TT is also supported in
part by JSPS Grant-in-Aid for Scientific Research No.20740132, and
by JSPS Grant-in-Aid for Creative Scientific Research No.\,19GS0219.

\noindent


\newcommand{\J}[4]{{\sl #1} {\bf #2} (#3) #4}
\newcommand{\andJ}[3]{{\bf #1} (#2) #3}
\newcommand{\AP}{Ann.\ Phys.\ (N.Y.)}
\newcommand{\MPL}{Mod.\ Phys.\ Lett.}
\newcommand{\NP}{Nucl.\ Phys.}
\newcommand{\PL}{Phys.\ Lett.}
\newcommand{\PR}{ Phys.\ Rev.}
\newcommand{\PRL}{Phys.\ Rev.\ Lett.}
\newcommand{\PTP}{Prog.\ Theor.\ Phys.}
\newcommand{\hep}[1]{{\tt hep-th/{#1}}}

\end{document}